\documentclass[%
aps,
prd,
preprint,
groupedaddress,
nofootinbib,
amsmath,
amssymb]{revtex4-1}

\usepackage{indentfirst}  
\usepackage{bm}    
\usepackage{graphicx}  
\usepackage{latexsym}
\usepackage{graphicx}
\usepackage{psfrag}
\usepackage{epsfig}
\usepackage{amsmath}
\usepackage{tocloft}
\usepackage{array}
\usepackage{setspace}
\usepackage{amssymb}
\usepackage{float}
\usepackage{multirow}
\usepackage{slashbox}
\usepackage{color}

\begin{document}

\title{Impulsive cylindrical gravitational wave: one possible radiative form emitted from cosmic strings and corresponding electromagnetic response}   



\author{H. Wen}
\affiliation{Department of Physics, Chongqing University, Chongqing 400044, P. R. China.}
\affiliation{College of Materials Science and Engineering, Chongqing University, Chongqing 400044, P.R. China}

\author{F.Y. Li}%
\email[]{cqufangyuli@hotmail.com}
\affiliation{Department of Physics, Chongqing University, Chongqing 400044, P. R. China.}
\affiliation{College of Materials Science and Engineering, Chongqing University, Chongqing 400044, P.R. China}

\author{Z.Y. Fang}%
\affiliation{Department of Physics, Chongqing University, Chongqing 400044, P. R. China.}
\affiliation{College of Materials Science and Engineering, Chongqing University, Chongqing 400044, P.R. China}

\author{A. Beckwith}%
\affiliation{Department of Physics, Chongqing University, Chongqing 400044, P. R. China.}
\affiliation{College of Materials Science and Engineering, Chongqing University, Chongqing 400044, P.R. China}



\begin{abstract}
\indent The cosmic strings(CSs) may be one type of important source of gravitational waves(\mbox{GWs}), and it has been intensively studied due to its special properties such as the cylindrical symmetry. The CSs would generate not only usual continuous \mbox{GW}, but also impulsive \mbox{GW} that brings more \mbox{concentrated} energy and consists of different GW components broadly covering low-, intermediate- and high-frequency bands simultaneously. These features might underlie interesting electromagnetic(EM) response to these GWs generated by the CSs.  In this paper, with novel results and effects, we firstly calculate the analytical solutions of perturbed EM fields caused by interaction between impulsive cylindrical \mbox{GWs} (would be one of possible forms emitted from CSs) and background
 celestial high magnetic fields or widespread cosmological background magnetic fields, by using rigorous \mbox{Einstein}-\mbox{Rosen} \mbox{metric} rather than the planar approximation usually applied. The results show that: perturbed EM fields are also in the impulsive form accordant to the GW pulse, and asymptotic behaviors of the perturbed EM fields are fully consistent with the asymptotic behaviors of the energy density, energy flux density and Riemann curvature tensor of corresponding impulsive cylindrical GWs. The analytical solutions naturally give rise to the accumulation effect(due to the synchro-propagation of perturbed EM fields and the GW pulse, because of their identical propagating velocities, i.e., speed of light), which is proportional to the term of  $\sqrt{distance}$. Based on this accumulation effect, in consideration of very widely existing background galactic-extragalactic magnetic fields in all galaxies and galaxy clusters, we for the first time predict potentially  observable effects in region of the Earth caused by the EM response to GWs from the CSs.

\begin{description}
\item[PACS numbers]
04.30.Nk,  04.25.Nx,  98.80.Cq,  04.80.Nn
\end{description}
\end{abstract}

\maketitle

\section{Introduction}
\label{intro}

\indent Over the past century, direct detection of gravitational wave(GW) has been regarded as one of
the most rigorous and ultimate tests of General Relativity, and has always been deemed as of
significant urgency and attracting extensive interest, by use of various observation schemes
aiming on multifarious sources.  Recently, detection of the B-mode polarization of cosmic microwave background has been reported\cite{B.mode.CMB}, and once this result obtains completely confirmed, it must be a great encouragement for such goal of GW detection.

\indent Other than usual GW origins, we specially focus on another important GW source, namely, the cosmic strings (CSs), a kind of axially symmetric cosmological body, which has been intensively researched \cite{Allen_arXiv9604033, Vilenkin_PLB47_1981,Caldwell_PRD3447_1992,Vachaspati_PhysRevD.31.3052,HOGAN_Nature_1984,
Damour_PRL3761_2000,Damour_PRD063510_2005,Leblond_PRD123519_2009,
Dufaux_PRD083518_2010,Berezinsky_PRD043004_2001,Copeland_JHEP013_2004,
Siemens.PRD.105001} in past decades, including issues related to impulsive GWs\cite{Podolsky.CQG.1401.2000,Podolsk.PhysRevD.81.124035, Gleiser.CQG.L141.1989,Slagter.CQG.463.2001,Hortacsu.CQG.2683,Steinbauer.CQG.2006} and continuous GWs\cite{Dubath.PRD.024001, Olmez_PhysRevD.81.104028, Patel.PRD.084032}. CSs are one-dimensional objects that can be formed in the early universe as the linear defects during
symmetry-breaking phase-transition\cite{Kleidis.2010.31,Hindmarsh_RepProgrPhys447_1995,Vilenkin_Cambridge_2000},
 so it represents the infinitely long line-source that would emit cylindrical \mbox{GWs}\cite{AnzhongWang.CQG.715,Gregory.PhysRevD.39.2108}. Because of  these particularities,  although the existence of the CSs has not been exactly concluded so far, \mbox{GWs} produced by the \mbox{CSs} already have attracted attentions from several efforts of observations by some main laboratories or projects,
 such as ground-based GW detectors\cite{Abbott_PRD062002_2009,Siemens_PhysRevLett.98.111101,Binetruy_PhysRevD.82.126007}
 and proposed space detector\cite{Cohen_CQG185012_2010,Callaghan_PhysRevLett.105.081602} in low- or intermediate-frequency bands.\\
 \indent Actually, the \mbox{GWs} generated by \mbox{CSs}, could have quite wide spectrum\cite{Vilenkin_PLB47_1981,HOGAN_Nature_1984,
 Vachaspati_PhysRevD.31.3052,Bennett_PhysRevLett.60.257,
 Caldwell_PRD3447_1992,Caldwell_PhysRevD.54.7146,Sarangi.PLB185.2002} even over $10^{10} Hz$\cite{Caldwell_PhysRevD.54.7146,Damour_PRL3761_2000,Olmez_PhysRevD.81.104028}.
Due to the cylindrical symmetrical property and the broad frequency range of these \mbox{GWs}, it's very interesting to consider the interaction between EM system and the cylindrical \mbox{GWs} from CSs, because the EM system could be quite suitable to reflect the particular characteristics of cylindrical \mbox{GWs} for many reasons:  the EM systems (natural or in laboratory) are widely existing (e.g. celestial and cosmological background magnetic fields); the GW and EM signal have identical propagating velocity, then leading to the spatial accumulation effect\cite{FYLi_PRD67_2003,FYLi_EPJC_2008,FYLi_PRD80_2009}; the EM system is generally sensitive to the \mbox{GWs} in a very wide frequency range (especially suitable to the impulsive form because the pulse comprises different GW components among broad frequency bands), and so on. \\

\indent In this paper we study the perturbed EM fields caused by an interaction between the EM system and the impulsive cylindrical \mbox{GWs} which could be emitted from the CSs and propagate through the background magnetic field\cite{DeLogi_PRD16_1977,Boccaletti_NuovoCim70_1970}; based on the rigorous \mbox{Einstein}-\mbox{Rosen} metric \cite{Einstein.JFI.43.1937,Rosen.PhyZSowjet.366.1937}(unlike usual planar approximation for weak \mbox{GWs}),
  analytical solutions of this perturbed EM
fields are obtained, by solving second order non homogeneous partial differential equation groups
(from electrodynamical equations in curved spacetime).\\
\indent Interestingly, our results show that the acquired solutions of perturbed EM fields are also in the form of a pulse, which is consistent to the impulsive cylindrical GW; and the asymptotic behavior of our solutions are in accordant to the asymptotic behavior of the energy-momentum tensor and the Riemann curvature tensor of the cylindrical GW pulse. This confluence greatly
supports the reasonability and self-consistence of acquired solutions.
 \\
\indent Due to identical velocities of the GW pulse and perturbed EM signals, the perturbed EM fields will be accumulated within the region of background magnetic fields, similarly to previous
research results\cite{FYLi_EPJC_2008,FYLi_PRD67_2003,FYLi_PRD80_2009}. Particularly, this accumulation effect is naturally reflected by our analytical solutions, and it's derived that the perturbed EM signals will be accumulated by the term of the square root of the propagating distance, i.e.   $\propto\sqrt{distance}$. Based on this accumulation effect, we first predict the possibly observable effects on the Earth(direct observable effect) or the indirectly observable effects(around magnetar), caused by EM response to the cylindrical impulsive GWs, in the background galactic-extragalactic magnetic fields ($\sim10^{-11}$ to $10^{-9}$ Tesla within 1Mpc\cite{galacticB.RevModPhys.74.775} in all galaxies
 and galaxy clusters) or strong magnetic surface fields of the magnetar\cite{Metzger.AstroPhysJourn.659.561.2007}($\sim 10^{11} Tesla$ or even higher). \\

It should be pointed out that, CSs produce not only usual continuous \mbox{GWs} \cite{Dubath.PRD.024001, Olmez_PhysRevD.81.104028, Patel.PRD.084032}, but also
impulsive \mbox{GWs} which have held a special fascination for researchers \cite{Podolsky.CQG.1401.2000,Podolsk.PhysRevD.81.124035, Gleiser.CQG.L141.1989, Slagter.CQG.463.2001,Hortacsu.CQG.2683,Steinbauer.CQG.2006} (e.g., the `Rosen'-pulse is believed to bring energy from the source of CSs\cite{Slagter.CQG.463.2001}).
  In this paper, we will specifically focus on the impulsive cylindrical \mbox{GWs}, and will discuss issues relevant to the continuous form in works done elsewhere. Some major reasons for this consideration include:\\
\indent (1)The impulsive \mbox{GWs} come with a very concentrated energy to give comparatively higher GW strength. In fact, this advantage is also beneficial to the detection by Adv-LIGO or LISA, eLISA (they may be very promising for \mbox{GWs} in the intermediate band($\nu\sim1Hz$ to $1000Hz$) and low frequency bands($10^{-6}$ to $10^{-2}Hz$), and it's possible to directly detect the continuous \mbox{GWs} from the CSs). However, the narrow width of GW pulse gives rise to greater proportion of energy distributed in the  high-frequency bands (e.g. GHz band), and it's already out of aimed frequency range of Adv-LIGO or LISA. However, the EM response could be suitable to these \mbox{GWs} with high-frequency components. \\
\indent (2)By Fourier decomposition, a pulse actually consists of different components of \mbox{GWs} among very wide
frequency range covering the low-, intermediate- and high-frequency signals simultaneously; these rich components make it particularly well suited to the EM response which is generally sensitive to broad frequency bands. \\
\indent (3)The exact metric of impulsive cylindrical GW underlying our calculation, has already been derived and developed in previous works
(by \mbox{Einstein}, \mbox{Rosen}, \mbox{Weber} and \mbox{Wheeler}\cite{Einstein.JFI.43.1937,Rosen.PhyZSowjet.366.1937,weber.1961.GRGW,Weber.RevModPhys.509.1957}), to provide a dependable and ready-made theoretical foundation. \\

The plan of this paper is as follows. In section II, the interaction between impulsive cylindrical
 \mbox{GWs} and background magnetic field is discussed.
 \mbox{In section III}, analytical solutions of the perturbed EM fields are calculated. In section IV, physical properties of the obtained solutions are in detail
 studied and demonstrated. In section V, EM response to the GWs in some celestial and cosmological conditions,
 and relevant potentially observable effects are discussed.
  In section VI, asymptotic physical behaviors of the perturbed EM fields are analyzed with
 comparisons to asymptotic behaviors of the GW pulse. In section VII, the conclusion and discussion are given, with both theoretical and observational perspectives for possible future     subsequent work along these lines.\\

\section{Interaction of the impulsive cylindrical GW with  background  magnetic field: a probable EM response to the GW}
\label{intro}

In Fig.1, the EM perturbation caused by cylindrical impulsive \mbox{GWs} within the background magnetic field is portrayed. Here, the CS (along $z$ axis) represents a line-source which produces \mbox{GWs} with cylindrical symmetry, and the impulsive \mbox{GWs} emitted from this \mbox{CS} propagate outwards in different directions, so we can chose one specific direction (the $x$ direction, perpendicular to the \mbox{CS}, see Fig.1) to focus on. In the interaction region near axis (symmetrical axis of the \mbox{CS}), a static
(or slowly varying quasi-static) magnetic field ${B^{_{(0)}}_{z}}$ is existing as interactive  background pointing to the $z$ direction. According to  electrodynamical equations in curved spacetime\cite{DeLogi_PRD16_1977,Boccaletti_NuovoCim70_1970}, these  cylindrical GW pulses will \mbox{perturb} this background magnetic field, and lead to perturbed EM fields (or in quantum language, signal photons) generated  within the region of background magnetic field;  then, the perturbed EM fields simultaneously and synchronously propagate in identical \mbox{direction} with the impulsive GWs along the $x$-axis.

\begin{figure}[!htbp]
        \centerline{\includegraphics[scale=0.65]{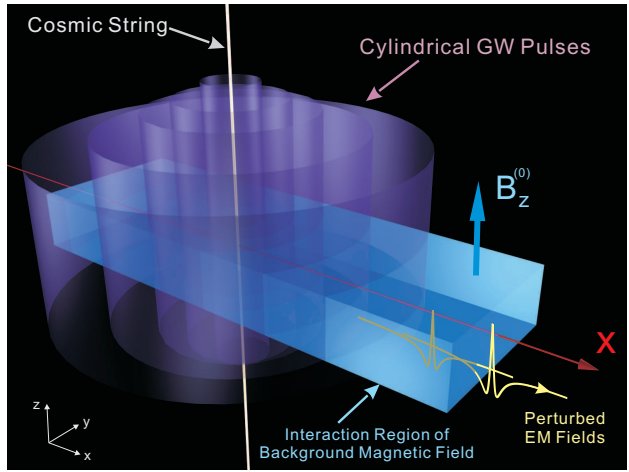}}
\begin{spacing}{1.2}
\caption{\footnotesize{\textbf{Interaction between cylindrical impulsive GWs and background magnetic field.} The cosmic string(alone $z$ axis) is emitting GW pulses propagating outwards perpendicularly to the CS, and we only focus on the EM perturbation in the $x$ direction specifically, the same hereinafter. The GW pulses will perturb the background magnetic field ${B^{_{(0)}}_{z}}$ (pointing to the $z$ axis) and produce the perturbed EM fields propagating in the $x$ direction. }}
\end{spacing}
\end{figure}

As aforementioned in section I, the cosmic strings could generate impulsive \mbox{GWs}\cite{Podolsky.CQG.1401.2000,Podolsk.PhysRevD.81.124035, Gleiser.CQG.L141.1989,Slagter.CQG.463.2001,Hortacsu.CQG.2683,Steinbauer.CQG.2006}
  with broad frequency bands\cite{Vilenkin_PLB47_1981,HOGAN_Nature_1984,
  Vachaspati_PhysRevD.31.3052,Bennett_PhysRevLett.60.257,Caldwell_PRD3447_1992,
  Caldwell_PhysRevD.54.7146,Damour_PRL3761_2000,Olmez_PhysRevD.81.104028}.
The key profiles for impulsive \mbox{GWs} are the pulse width ``a'', the amplitude ``A'' and its
specific metric. Here, for the convenience and clearness, we select the \mbox{Einstein}-\mbox{Rosen} metric
 to describe the cylindrical impulsive GWs. This well known metric initially derived by \mbox{Einstein} and \mbox{Rosen} based on General
 Relativity\cite{Einstein.JFI.43.1937,Rosen.PhyZSowjet.366.1937,Rosen.GRG.429.1993},  has been widely researched, such as  pertinent
  issues of the energy-momentum pseudo-tensor\cite{Rosen.GRG.429.1993,Rosen.HeivPhysActa.171.1956,Rosen.PhysRev.291.1958,Weber.RevModPhys.509.1957,FYLi.ActaPhysSin.321.1997}; its \mbox{concise} and \mbox{succinct} form could be advantageous to reveal the impulsive and cylindrical symmetrical properties of \mbox{GWs}. Using cylindrical polar coordinates $(\rho, \varphi, z)$ and time $t$,
 the \mbox{Einstein}-\mbox{Rosen} metric of the impulsive cylindrical GW can be written as\cite{Einstein.JFI.43.1937, Rosen.PhyZSowjet.366.1937, weber.1961.GRGW}
  ($c = 1$ in natural unit):
\begin{eqnarray}
\label{eq1}
~~~ds^2 &=& e^{2(\gamma-\psi)}(dt^2-d\rho^2)-e^{-2\psi}\rho^2d\varphi^2-e^{2\psi}dz^2,
\end{eqnarray}
then, the contravariant components of the metric \mbox{tensor} $g^{\mu\nu}$ are:
\begin{eqnarray}
\label{eq2}
~~~~~~~~~~g^{00}&=&e^{2(\psi-\gamma)}, ~g^{11}=-e^{2(\psi-\gamma)},\nonumber\\
g^{22}&=&-e^{2\psi},   ~~~~g^{33}=-e^{-2\psi}.
\end{eqnarray}
and we have
\begin{eqnarray}
\label{eq3}
~~~~~~~~~~~\sqrt{-g}=e^{2(\gamma-\psi)}
\end{eqnarray}
\noindent here, the $\psi$ and $\gamma$ are functions with respect to the distance $\rho$ (will be denoted as `$x$' in the coordinates after this section) and the time $t$\cite{Einstein.JFI.43.1937, Rosen.PhyZSowjet.366.1937, weber.1961.GRGW,  Weber.RevModPhys.509.1957, Rosen.GRG.429.1993}, namely:
\begin{equation}
\label{eq4}
\psi(\rho,t)=A\{\frac{1}{[(a-it)^2+\rho^2]^{\frac{1}{2}}}+{\frac{1}{[(a+it)^2+\rho^2]^{\frac{1}{2}}}}\}
\end{equation}
\begin{eqnarray}
\label{eq5}
\gamma(\rho,t)&=&\frac{A^2}{2}\{\frac{1}{a^2}-\frac{\rho^2}{[(a-it)^2+\rho^2]^2}-\frac{\rho^2}{[(a+it)^2+\rho^2]^2}\nonumber\\
&-&\frac{t^2+a^2-\rho^2}{a^2[t^4+2t^2(a^2-\rho^2)+(a^2+\rho^2)^2]^{\frac{1}{2}}}\},
\end{eqnarray}
\noindent where $A$ and  $a$ are corresponding to the amplitude and width of the cylindrical impulsive GW, respectively.\\

\section{The perturbed EM fields produced by the impulsive cylindrical GW in the background magnetic field}
\indent When the cylindrical impulsive GW described in section II (Eqs.(1) to (5))
propagates through the interaction region with background  magnetic
field ${B^{_{(0)}}_{z}}$, the perturbed EM fields will be generated. In this section, we will formulate a detailed calculation on the exact forms of the perturbed EM fields. Notice that, the ``exact'' here means we utilize the rigorous metric of cylindrical impulsive GW(Eq.1) which keeps the cylindrical form,  rather than  the planar approximation usually used for weak fields by linearized \mbox{Einstein} equation. Nonetheless, the cross-section of interaction between background magnetic field and the GW pulse, is still very small, so the consideration of perturbation theory is reasonably suited to handle this case, and as commonly accepted process, some high order \mbox{infinitesimals} can be ignored. However, the manipulating without using the perturbation theory and without   any sort of approximation to seek absolutely strict results, is also very interesting topic, and we would  investigate such issues in other works. Therefore, the total EM field \mbox{tensor} $F_{\alpha\beta}$ can be
expressed as two parts: the background static magnetic field ${F^{_{(0)}}_{\alpha\beta}}$,
 and  the perturbed EM fields ${F^{_{(1)}}_{\alpha\beta}}$ caused by the incoming impulsive GW; Because of the cylindrical symmetry, it is always possible to describe the EM perturbation effect at the plane of $y=0$ (i.e., the x-z plane,  means by use of a local Cartesian coordinate system, see Fig.1, and the `x' here substitutes for the distance $\rho$ in Eqs(4) and (5)), then $F_{\alpha\beta}$ can be written as:\\
\begin{eqnarray}
\label{eq6}
&~&F_{\alpha \beta } ={F^{_{(0)}}_{\alpha\beta}}+ {F^{_{(1)}}_{\alpha\beta}}\nonumber\\
~\nonumber\\
&=&\left( {{\begin{array}{*{20}c}
 0  &  {E^{_{(1)}}_x}  &{E^{_{(1)}}_y} & &{E^{_{(1)}}_z}&  \\
 ~\nonumber\\
-{E^{_{(1)}}_x} & 0  &  (-{B^{_{(0)}}_{z}}-{B^{_{(1)}}_z}) & &{B^{_{(1)}}_y} & \\
~\nonumber\\
-{E^{_{(1)}}_y}  & ({B^{_{(0)}}_{z}}+{B^{_{(1)}}_z})  & 0  & -&{B^{_{(1)}}_x}&  \\
~\nonumber\\
-{E^{_{(1)}}_z} & -{B^{_{(1)}}_y}   &  {B^{_{(1)}}_x}    &  &0& \\
\end{array} }} \right)\nonumber\\
\end{eqnarray}
~\\
Then, using the electrodynamical equations in curved spacetime:\\
\begin{eqnarray}
\label{eq7}
\nabla_{\nu}F^{\mu\nu}&=&\frac{1}{\sqrt{-g}}\frac{\partial}{\partial x^{\nu}}[\sqrt{-g}g^{\mu\alpha}g^{\nu\beta} ({F^{_{(0)}}_{\alpha\beta}}+ {F^{_{(1)}}_{\alpha\beta}} )   ]=0,\nonumber\\
~\nonumber\\
\nabla_{\alpha}F_{\mu\nu}&+&\nabla_{\nu}F_{\alpha\mu}+\nabla_{\mu}F_{\nu\alpha} = 0,\nonumber\\
~\nonumber\\
where&,&   {F^{_{(0)}}_{12}} =- {F^{_{(0)}}_{21}} =-{B^{_{(0)}}_{z}}=- {B^{_{(0)}}}
\end{eqnarray}
~\\
and together with Eqs.(\ref{eq1}) to (\ref{eq6}), we have:\\
\begin{eqnarray}
\label{eq6a}
\mu=0 &\Rightarrow& ~~ 2(\gamma_x-\psi_x) {E^{_{(1)}}_x}-\frac{\partial{E^{_{(1)}}_x}}{\partial x}=0,\\
~\nonumber\\
\mu=1 &\Rightarrow& ~~ 2(\psi_t-\gamma_t~) {E^{_{(1)}}_x}+\frac{\partial{E^{_{(1)}}_x}}{\partial t}=0,\\
~\nonumber\\
\mu=2 &\Rightarrow& ~~ 2\psi_t{E^{_{(1)}}_y}+ \frac{\partial{E^{_{(1)}}_y}}{\partial t}\nonumber\\
~\nonumber\\
&+&2\psi_x( {B^{_{(0)}}} +{B^{_{(1)}}_z})+\frac{\partial{B^{_{(1)}}_z}}{\partial x}=0,\\
~\nonumber\\
\mu=3 &\Rightarrow& ~~ 2\psi_t{E^{_{(1)}}_z}+ \frac{\partial{E^{_{(1)}}_z}}{\partial t}-2\psi_x{B^{_{(1)}}_y}+\frac{\partial{B^{_{(1)}}_y}}{\partial x}=0,\nonumber\\
\\
\frac{\partial {B^{_{(1)}}_x} }{\partial x}&=&0,~~ \frac{\partial {B^{_{(1)}}_x} }{\partial t}=0, \nonumber\\
~\nonumber\\
\frac{\partial{E^{_{(1)}}_z}}{\partial x}&=&\frac{\partial{B^{_{(1)}}_y}}{\partial t}, ~~\frac{\partial{E^{_{(1)}}_y}}{\partial x}=-\frac{\partial{B^{_{(1)}}_z}}{\partial t}.
\end{eqnarray}
~\\
Here, $\gamma_{x}$, $\psi_{x}$, $\psi_{t}$ and $\gamma_{t}$ stand for  $\frac{\partial\gamma}{\partial x}$, $\frac{\partial\psi}{\partial x}$, $\frac{\partial\psi}{\partial t}$ and $\frac{\partial\gamma}{\partial t}$,  similarly hereinafter. So, by omitting second- and higher- order infinitesimal terms, it gives:\\
\begin{eqnarray}
\label{partialEqu}
\frac{\partial^2{E^{_{(1)}}_y}}{\partial x^2}-\frac{\partial^2{E^{_{(1)}}_y}}{\partial t^2}&=&2\psi_{xt} {B^{_{(0)}}} ,\\
~\nonumber\\
\frac{\partial^2{B^{_{(1)}}_z}}{\partial x^2}-\frac{\partial^2{B^{_{(1)}}_z}}{\partial t^2}&=&-2\psi_{xx} {B^{_{(0)}}} ,
\end{eqnarray}
~\\
Note that with Eqs.(8) to (12) and the initial conditions, we have:\\
\begin{eqnarray}
\label{partialEquBound}
{E^{_{(1)}}_y}|_{_{t=0}}&=&0,~~\frac{\partial{E^{_{(1)}}_y}}{\partial t}|_{_{t=0}}=-2\psi_{x}|_{_{t=0}}\cdot {B^{_{(0)}}} , \nonumber\\
~\nonumber\\
{B^{_{(1)}}_z}|_{_{t=0}}&=&0, ~~\frac{\partial{B^{_{(1)}}_z}}{\partial t}|_{_{t=0}}=0.
\end{eqnarray}
~\\
The other components, i.e., ${E^{_{(1)}}_x}$, $ {B^{_{(1)}}_x} $, ${E^{_{(1)}}_z}$, ${B^{_{(1)}}_y}$, have only null solutions. Non-vanishing EM components ${E^{_{(1)}}_y}$ and $ {B^{_{(1)}}_z} $ are functions of $x$ and $t$.  To obtain their solutions, we need to solve the group of second order non homogeneous partial
 differential equations in Eqs.(\ref{partialEqu}) to (\ref{partialEquBound}),
and utilizing the d'Alembert Formula, we can express the solutions in analytical way\cite{Tikhonov}:\\
\begin{eqnarray}
\label{EbeforeIntegral}
{E^{_{(1)}}_y}&=&\frac{1}{2}\int_{x-t}^{x+t}H(\xi)d\xi \nonumber\\
~\nonumber\\
&+&\frac{1}{2}\int_{0}^{t}\int_{x-(t-\tau)}^{x+(t-\tau)}F(\xi, \tau)d\xi d\tau,\\
~\nonumber\\
~\nonumber\\
{B^{_{(1)}}_z}&=& \frac{1}{2}\int_{0}^{t}\int_{x-(t-\tau)}^{x+(t-\tau)}G(\xi, \tau)d\xi d\tau,
\end{eqnarray}
where,
\begin{eqnarray}
\label{HbeforeIntegral}
&~&H(\xi)=-2\psi_{\xi}|_{_{t=0}}\cdot {B^{_{(0)}}} ,\nonumber\\
~\nonumber\\
&~&F(\xi, \tau)=-2\psi_{\xi\tau}  {B^{_{(0)}}} , \nonumber\\
~\nonumber\\
&~&G(\xi, \tau)= 2\psi_{\xi\xi}  {B^{_{(0)}}} .
\end{eqnarray}
~\\
By integral from Eq.(\ref{EbeforeIntegral}), one gives:\\
\begin{eqnarray}
\label{eq18}
&~&\frac{1}{2}\int_{x-t}^{x+t}H(\xi)d\xi = -1\int_{x-t}^{x+t}\psi_{\xi}|_{_{t=0}}\cdot {B^{_{(0)}}} d\xi \nonumber\\
~\nonumber\\
&=& 2A {B^{_{(0)}}} \{\frac{1}{[(x-t)^2+a^2]^{\frac{1}{2}}}-\frac{1}{[(x+t)^2+a^2]^{\frac{1}{2}}}\}.
\end{eqnarray}\\
and similarly,
\begin{eqnarray}
\label{eq19}
&~&\frac{1}{2}\int_{0}^{t}\int_{x-(t-\tau)}^{x+(t-\tau)}F(\xi, \tau)d\xi d\tau \nonumber\\
&=&-A {B^{_{(0)}}} \{\int_{0}^t\frac{\tau+ia}{[(x+t-\tau)^2+(a-it)^2]^{\frac{3}{2}}}d\tau\nonumber\\
&+&\int_{0}^t\frac{\tau-ia}{[(x+t-\tau)^2+(a+it)^2]^{\frac{3}{2}}}d\tau \nonumber\\
&-&\int_{0}^t\frac{\tau+ia}{[(x-t+\tau)^2+(a-it)^2]^{\frac{3}{2}}}d\tau\nonumber\\
&-&\int_{0}^t\frac{\tau-ia}{[(x-t+\tau)^2+(a+it)^2]^{\frac{3}{2}}}d\tau\},
\end{eqnarray}
and
\begin{eqnarray}
\label{eq19a}
&~&\frac{1}{2}\int_{0}^{t}\int_{x-(t-\tau)}^{x+(t-\tau)}G(\xi, \tau)d\xi d\tau \nonumber\\
&=&-A {B^{_{(0)}}} \{\int_{0}^t\frac{x+t-\tau}{[(x+t-\tau)^2+(a-i\tau)^2]^{\frac{3}{2}}}d\tau\nonumber\\
&+&\int_{0}^t\frac{x+t-\tau}{[(x+t-\tau)^2+(a+i\tau)^2]^{\frac{3}{2}}}d\tau \nonumber\\
&-&\int_{0}^t\frac{x-t+\tau}{[(x-t+\tau)^2+(a-i\tau)^2]^{\frac{3}{2}}}d\tau\nonumber\\
&-&\int_{0}^t\frac{x-t+\tau}{[(x-t+\tau)^2+(a+i\tau)^2]^{\frac{3}{2}}}d\tau\}
\end{eqnarray}
After lengthy calculation of \mbox{Eqs.} (\ref{eq19}) and (\ref{eq19a}), with combination of the Eq. (\ref{eq18}),
the concrete form of \mbox{electric} component of the perturbed EM fields can be deduced:

\begin{widetext}
\begin{eqnarray}
\label{E}
{E^{_{(1)}}_y}(x, t)  &=&  2A {B^{_{(0)}}} \{\frac{1}{[(x-t)^2+a^2]^{\frac{1}{2}}}-\frac{1}{[(x+t)^2+a^2]^{\frac{1}{2}}}\nonumber\\
&+& \frac{ a (x+t)\sin(\frac{1}{2}\theta_1)-a^2\cos(\frac{1}{2}\theta_1)}   { [(x+t)^2+a^2][x^4+2x^2(a^2-t^2)+(a^2+t^2)^2]^{\frac{1}{4}}    }  \nonumber\\
   &+&\frac{a^2}{[(x+t)^2+a^2]^{\frac{3}{2}}    }-\frac{a^2}{[(x-t)^2+a^2]^{\frac{3}{2}}  }+ \frac{a(x-t)\sin[\frac{1}{2}\theta_1]+a^2\cos[\frac{1}{2}\theta_1]} {[(x-t)^2+a^2][x^4+2x^2(a^2-t^2)+(a^2+t^2)^2]^{\frac{1}{4}}} ~\}\nonumber\\
 + A {B^{_{(0)}}} &\{& \frac{\cos[2\theta_2+\frac{1}{2}\theta_1]   }{[x^4+2x^2(a^2-t^2)+(a^2+t^2)^2]^{\frac{1}{4}}}  -\frac{2\cos[2\theta_2]}{[(x-t)^2+a^2]^{\frac{1}{2}}} \nonumber\\
 &+&\frac{[x^4+2x^2(a^2-t^2)+(a^2+t^2)^2]^{\frac{1}{4}}  }{(x-t)^2+a^2    }\cos[2\theta_2-\frac{1}{2}\theta_1]  \nonumber\\
&-&\frac{\cos[2\theta_3-\frac{1}{2}\theta_1]}{[x^4+2x^2(a^2-t^2)+(a^2+t^2)^2]^{\frac{1}{4}}}+\frac{2\cos(2\theta_3)}{[(x+t)^2+a^2]^{\frac{1}{2}}} \nonumber\\
&-&\frac{[x^4+2x^2(a^2-t^2)+(a^2+t^2)^2]^{\frac{1}{4}}}{(x+t)^2+a^2}\cos[2\theta_3+\frac{1}{2}\theta_1]\} \nonumber\\
\end{eqnarray}

The same procedure, from Eqs.(14), (15), (17) and (18), will lead to the following , i.e. the concrete form of magnetic component of the perturbed EM fields can be derived to read as:

\begin{eqnarray}
\label{H}
{B^{_{(1)}}_z}(x, t) &=& -2A {B^{_{(0)}}} \{\frac{(x+t)^2\cos(\frac{\theta_1}{2})+(x+t)a\sin(\frac{\theta_1}{2})}{[(x+t)^2+a^2][x^4+2x^2(a^2-t^2)+(a^2+t^2)^2]^{\frac{1}{4}}}
-\frac{(x+t)^2}{[(x+t)^2+a^2]^{\frac{3}{2}}}\nonumber\\
&+&\frac{(x-t)^2\cos(\frac{\theta_1}{2})-(x-t)a\sin(\frac{\theta_1}{2})}{[(x-t)^2+a^2][x^4+2x^2(a^2-t^2)+(a^2+t^2)^2]^{\frac{1}{4}}}
-\frac{(x-t)^2}{[(x-t)^2+a^2]^{\frac{3}{2}}}\}\nonumber\\
+A {B^{_{(0)}}} &\{&\frac{\cos(2\theta_3-\frac{1}{2}\theta_1)}{[x^4+2x^2(a^2-t^2)+(a^2+t^2)^2]^{\frac{1}{4}}}-\frac{2\cos(2\theta_3)}{[(x+t)^2+a^2]^{\frac{1}{2}}}
 \nonumber\\
 &+&\frac{[x^4+2x^2(a^2-t^2)+(a^2+t^2)^2]^{\frac{1}{4}}}{(x+t)^2+a^2}\cos(2\theta_3+\frac{1}{2}\theta_1) \nonumber\\
&+&\frac{\cos(2\theta_2+\frac{1}{2}\theta_1)}{[x^4+2x^2(a^2-t^2)+(a^2+t^2)^2]^{\frac{1}{4}}}-\frac{2\cos(2\theta_2)}{[(x-t)^2+a^2]^{\frac{1}{2}}}
 \nonumber\\
 &+&\frac{[x^4+2x^2(a^2-t^2)+(a^2+t^2)^2]^{\frac{1}{4}}}{(x-t)^2+a^2}\cos(2\theta_2-\frac{1}{2}\theta_1) ~\} \nonumber\\
\end{eqnarray}
where, the three angles used in Eq.(23) read as given below in Eq. (24), namely:\\
\begin{eqnarray}
~\theta_1= \arctan\frac{2at}{x^2+a^2-t^2}, ~\theta_2=\arctan\frac{a}{x-t}, ~\theta_3=\arctan\frac{a}{x+t}.
\end{eqnarray}
\end{widetext}
\begin{figure}[!htbp]
\centerline{\includegraphics[scale=0.7]{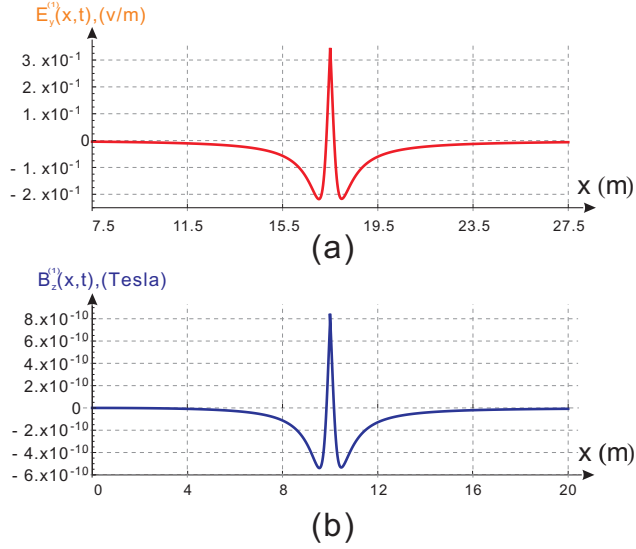}}
\begin{spacing}{1.2}
\caption{\footnotesize{\textbf{Typical examples of electric and magnetic components of perturbed impulsive EM fields in the region near axis.}
According to Eq.(\ref{E}), Fig.2(a) shows a representative electric component ${E^{_{(1)}}_y}(x, t)$ of impulsive EM field produced by
interaction between cylindrical GW pulse(width $a=0.4 m$, amplitude $h\sim 10^{-21}$, here we denote `$h$' instead of `A' as the amplitude in the SI units, similarly hereinafter) and
higher background magnetic field($10^{11} Tesla$, could be generated from celestial bodies\cite{Metzger.AstroPhysJourn.659.561.2007}, the same  hereinafter). Here we assume $h\sim 10^{-21}$ (similarly in later figures), and the detailed reason for this choice may be found in Section V.  In the same way, according to Eq.(\ref{H}), Fig.2(b) presents a typical magnetic component ${B^{_{(1)}}_z}(x, t)$  of perturbed EM fields(with GW width $a=0.4 m$, $h\sim 10^{-21}$) in the region of near axis. Both electric and magnetic components of perturbed impulsive EM fields are in the form of pulse, which are
consistent with respect to the GW pulses. This figure is plotted under SI units.}}
\end{spacing}
\end{figure}

The analytical solutions given above of the electric component ${E^{_{(1)}}_y}(x, t)$
and the magnetic component ${B^{_{(1)}}_z}(x, t)$ of the perturbed EM fields, give concrete description of
the interaction between the GW pulse and background  magnetic field. Here, ${E^{_{(1)}}_y}(x, t)$
and ${B^{_{(1)}}_z}(x, t)$ are both functions of time $t$ and coordinate $x$, with parameters ``$A$'' (amplitude
of the GW), ``$a$'' (width of the GW pulse) and $ {B^{_{(0)}}} $ (background magnetic field). So, these
solutions contain essential information inherited from their GW source(e.g. cosmic string) and EM system(e.g., celestial or galactic-extragalactic  background magnetic fields).\\
\indent It is important to note that the analytical solutions Eq.(\ref{E}) and (\ref{H}) of the perturbed EM field are also in the form of a pulse (see Fig.2). Both Eq.(22) and Eq.(23)  are consistent with regards to the impulsive GWs. In Fig.2(a),  this fact is explicitly revealed in that  at a specific time $t$, the waveform
of the electric field ${E^{_{(1)}}_y}(x, t)$  exhibits  a peak in magnitude.  A similar situation also occurs as exhibited in Fig.2(b) for  the magnetic component of the perturbed EM fields.\\

\indent On account of the cylindrical symmetry of the GW pulse, its source should be some  one-dimensionally distributed object of very large scale, and according to current cosmological observations or theories, cosmic strings would almost certainly be the best candidate. Correspondingly, from the solutions, these impulsive peaks of perturbed EM fields are found to propagate outwards  from the symmetrical axis of the CS, with their wavefronts also following a cylindrical manner (see Fig.3), and meanwhile  their levels gradually increase during this propagation process (due to the accumulation effect, will be then discussed later in the paper).\\

The analytical solutions Eqs.(\ref{E}) and (\ref{H}) which are in the impulsive manner, bring us important
key information and show the special nature of the perturbed EM fields,
as well as underlying various physical properties, potentially observable effects and asymptotic behaviors. These aspects will be consequently analyzed and discussed in full  in the  following sections.

\begin{figure}[!htbp]
\centerline{\includegraphics[scale=0.54]{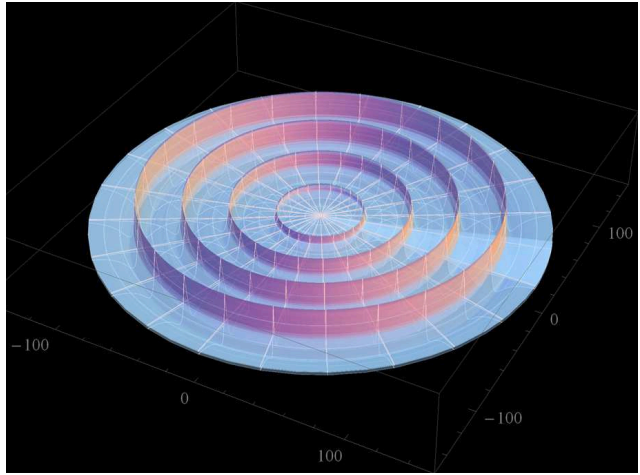}}
\begin{spacing}{1.2}
\caption{\footnotesize{\textbf{Wavefronts of the perturbed EM fields in background  magnetic field.}
This plot generated above is based upon Eq.(\ref{E}), and it demonstrates the cylindrical wavefronts of the EM field as
perturbed by the cylindrical GW pulses.
}}
\end{spacing}
\end{figure}

\section{Physical properties of analytical solutions of the perturbed EM fields}

With the  analytical solutions of the perturbed EM fields  as represented in the above section, some of their interesting properties may be studied in detail. As an example, what is the relationship between the given width-amplitude of perturbed EM fields and the width-amplitude of the GW pulse?  Secondly, is there any accumulation effect (consistent with previous work in the literature on the
 EM respond to GWs) of the interaction between the GW pulses and the background magnetic field since the GWs and the perturbed EM fields share the same velocity(speed of light)?  In addition we also ask what is the spectrum of the amplitude of the perturbed EM field and how it is related to
 the parameters of the corresponding GW pulse? For convenience, we use SI units in this section and section V. The details are as follows:\\

(1)\textbf{The relationship between parameters of the \mbox{GW} pulse and the impulsive features of the perturbed EM fields.} \\
\indent From Eq.(\ref{E}), it's simply deduced that the amplitude $h$ (here, `h' is in place of `A' as the amplitude in SI units) of the GW pulse and the background magnetic field $ {B^{{(0)}}} $,  contribute linear factors for the perturbed EM field which we designate in this paper as
${E^{_{(1)}}_y}(x, t)$. So the ${E^{_{(1)}}_y}(x, t)$ varies according to
 $h$ proportionally (see Fig.4(b)). However, the width `$a$' of a GW pulse plays a more complex
role in Eq.(\ref{E}), but we find that width 'a' is still positively correlated to the width of  the perturbed EM fields, i.e., a smaller width of the GW pulse leads to a smaller width of perturbed EM field
 (see Fig.4(a)), and the EM pulse with a narrower peak is found to have larger strength and more concentrated energy.
\begin{figure}[!htbp]
\centerline{\includegraphics[scale=0.73]{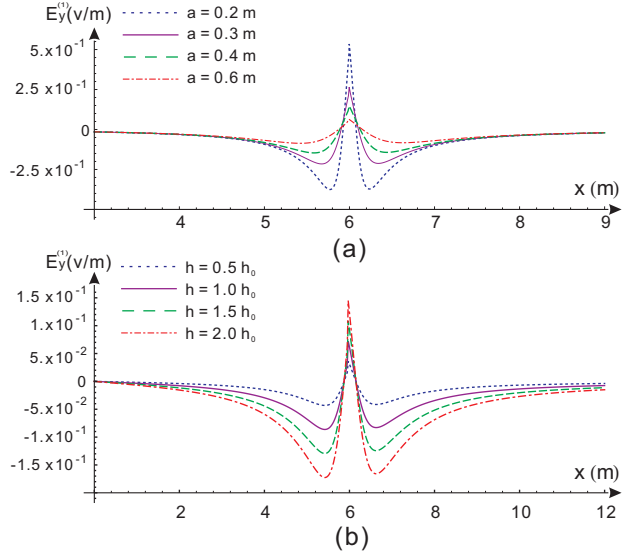}}
\begin{spacing}{1.2}
\caption{\footnotesize{\textbf{EM fields perturbed by the GW pulses with different widths and amplitudes.}
Here, the Fig.4(a) exhibits the electric components of EM fields perturbed by the GW pulses with different pulse widths(from $a=0.2m$ to $0.6m$, given the same
amplitude $\sim10^{-21}$ in the region near axis). It's indicated that a larger width of the GW pulse gives rise to a larger width of the corresponding perturbed EM pulse, but also results in a more flat waveform with less concentrated energy.  Fig.4(b) reveals the impact from different
amplitudes($h=0.5$ to $2~h_0$, $h_0\sim10^{-21}$) of the GW pulses with the same width, and explicitly we find that higher amplitudes of the GW pulse
lead to a corresponding greater amplitude of the perturbed EM field.  Note that the growth of the amplitude of GW does not influence the width of the perturbed EM pulse.}}
\end{spacing}
\end{figure}
~\\
(2)\textbf{Propagating velocity of perturbed EM pulses.} \\
\indent The information about propagating velocity of
the GW as the speed of light, is naturally included in the definition of the  metric(Eqs.(1) to (5)).
The same, EM pulses caused by the GW pulse, also are propagating at the speed of light due to EM theory in free space (also in curve space-time)\cite{DeLogi_PRD16_1977, Boccaletti_NuovoCim70_1970}. For intuitive representation,
 we exhibit this property in Fig.5, which illustrates the exact given field contours of ${E^{_{(1)}}_y}(x, t)$ at different time from $t=0$ to $t=\frac{20}{c}$ second.\\

\begin{figure}[!htbp]
\centerline{\includegraphics[scale=0.66]{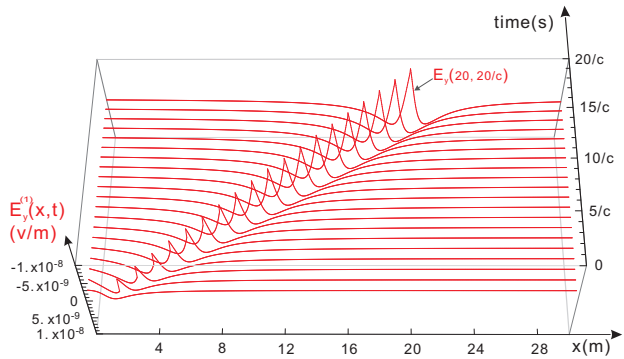}}
\begin{spacing}{1.2}
\caption{\footnotesize{\textbf{Perturbed EM field propagates at the speed of light.} This plot has a visual display of the propagating
process of a typical perturbed EM pulse from $t=0$ to $t=\frac{20}{c}$ second, which is
linearly moving from x = 0 m to x = 20 m during this period. I.e., the speed of propagating of the perturbed EM pulse is $20/\frac{20}{c}=c$, as the same as speed of light.}}
\end{spacing}
\end{figure}
(3)\textbf{Accumulation effect due to the identical propagating velocity of EM  pulses and GW pulse.} \\
\indent Mentioned above, we have that when the perturbed EM fields propagate at the speed of light
synchronously with the GW pulse, then, the perturbed EM fields caused by interaction between GW pulse and background
magnetic field will accumulate.
 So, in the region with a given background
 magnetic field, the strength of the perturbed EM pulse will rise (see Fig.6) gradually until it leaves the \mbox{boundary} of the background
 magnetic field.  Except the reason of their synchronous propagation, the accumulation should also be determined by
  another fact, that, the energy flux of impulsive cylindrical GW will decay by $1/distance$ (or $\sim1/x$ on light-cone, see Eq.(33), so the strength of impulsive cylindrical GW will decay by $1/\sqrt{distance}$), and their composite accumulation \mbox{effect} is proportional to $\sqrt{x}$ (see Eq.(\ref{Eaccumu})). In Fig.6, we can also find that the accumulation effect is conspicuous, and for diverse cases of perturbed EM fields with different  parameters of width of the GW pulses (see Fig.6), this phenomenon always appears generally, and in Fig.7, the contours  of perturbed EM fields(electric component)in different positions, also explicitly demonstrates the accumulated perturbed impulsive EM fields, during their propagating away from the GW source.\\
\begin{figure*}[!htbp]
\centerline{\includegraphics[scale=0.8]{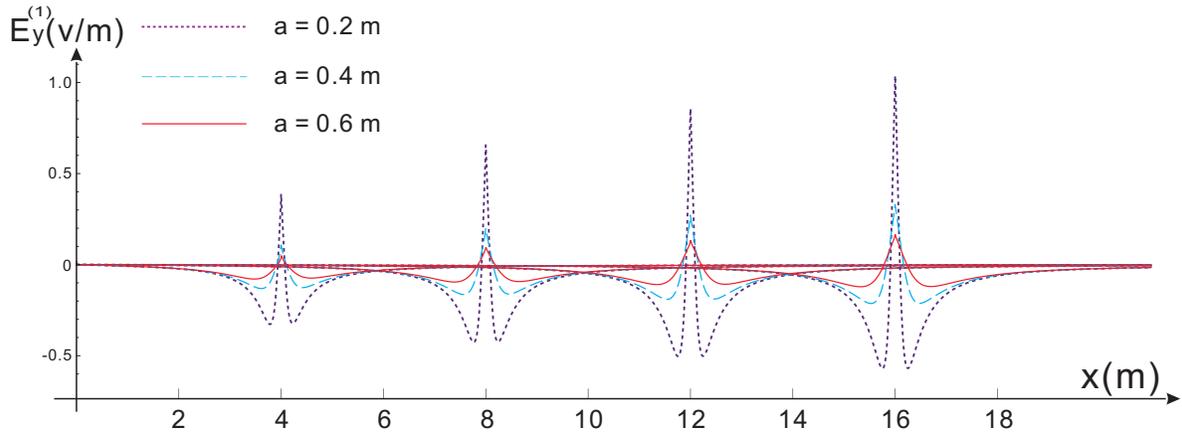}}
\begin{spacing}{1.2}
\caption{\footnotesize{\textbf{Accumulation effect of the perturbed EM fields.} When the impulsive EM fields propagate from $x=0$, at different positions of $x=4, 8, 12, 16m$ in the region near axis
(at different times, t, amplitude $h\sim 10^{-21}$, background magnetic field $\sim 10^{11}~Tesla$), the amplitude of the EM fields will then increase, because the perturbed EM fields propagate synchronously with the
GW pulse with the identical velocity of
the speed of light, then the interaction between the GW pulse and background magnetic field will
be finally accumulated. Here, this figure shows that this accumulation phenomenon commonly  appears in all cases with different widths from $0.2$ to $0.6m$ of the GW pulses.}}
\end{spacing}
\end{figure*}

\begin{figure*}[!htbp]
\centerline{\includegraphics[scale=0.5]{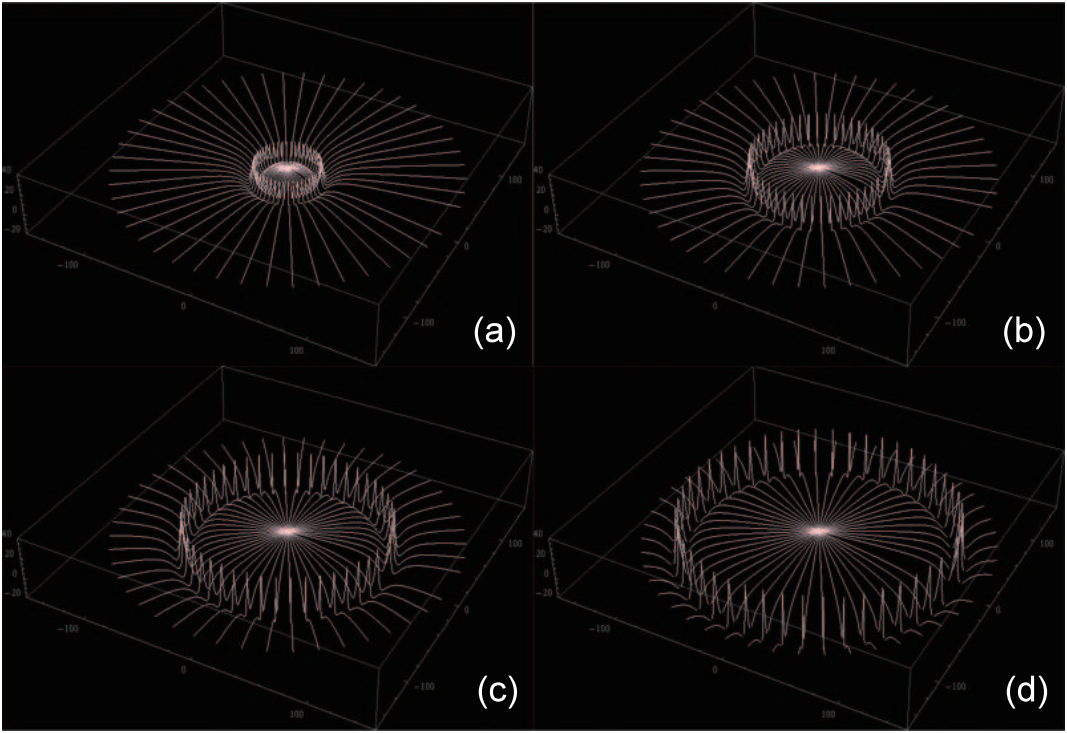}}
\begin{spacing}{1.2}
\caption{\footnotesize{\textbf{Contours of perturbed EM fields(electric component) in different positions (30m, 60m, 90m and 120m in sub-figures(a), (b), (c) and (d) respectively) away from the cosmic string, pulse width: 5m.} The accumulation effect is apparent during the propagation of the perturbed EM pulses outward from the symmetric axis.}}
\end{spacing}
\end{figure*}

(4)\textbf{Amplitude spectrums of perturbed EM fields influenced by amplitude of the GW pulse. }\\
\indent Mentioned above, the perturbed impulsive EM field is comprised of  components with very different frequencies among a  wide band. Shown in Fig.8, the Fourier transform of field ${E^{_{(1)}}_y}(x, t)$ in
 frequency domain, illustrates the distribution of the amplitude spectrum. Although this spectrum decreases as it approaches high-frequency range, it still remains available level in the area of GHz band. Overall, it is  indicated that (see Fig.8), the level of spectrum is proportional to the amplitude of the GW pulse which causes the perturbed EM fields, commonly among entire frequency bands.

\begin{figure}[!htbp]
\centerline{\includegraphics[scale=0.75]{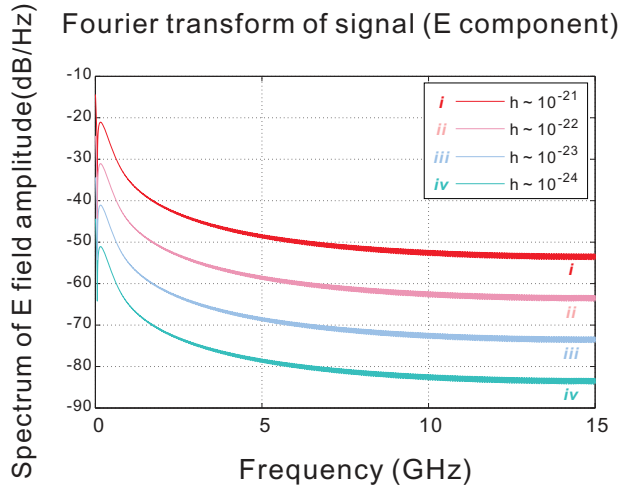}}
\begin{spacing}{1.2}
\caption{\footnotesize{\textbf{Amplitude spectrums of perturbed EM fields caused by the GW pulses with different amplitudes.}
This figure demonstrates the amplitude spectrums of perturbed EM fields (electric component) caused by the GW pulses
with different amplitudes from $\sim10^{-21}$ to $\sim10^{-24}$ (with identical width of $a=0.5m$). For all cases, the
spectrums will decay as the frequency goes higher, but still remain at a considerable level even over GHz band. It reflects the proportional
relationship between the overall level of spectrum among all frequency region and the amplitude of GW pulse. The unit ``dB'' here means $10\times log_{10}( )$.}}
\end{spacing}
\end{figure}
~\\
(5)\textbf{Amplitude spectrums of perturbed EM \mbox{fields} influenced by the width of GW pulse.} \\
\indent In contrast to what happens  to GW amplitude $h$, the width of GW pulse acts so as to impact
 the distribution of amplitude spectrums of the perturbed EM fields, apparently in a nonlinear
manner (see Fig.9). The narrow width of the GW pulse will lead to a rich spectrum in the high-frequency region, such as GHz band, as exemplified by the upper-left subplot of Fig.9, where
 the width of GW pulse is 0.1 meter, and hence the sum of energy of the spectral components
 from 1GHz to 9.9GHz is approximately $24.5\%$ out of the total energy. Once the width rises, as
shown in the other three subplots in Fig.9, gradually, the amplitude spectrums will decrease in
the high-frequency domain. This property elucidates why the smaller width of the GW pulse
would more likely cause stronger effect of EM response especially in the high-frequency
band, and this phenomenon is generally appearing in the wide continuous parameter range (see Fig.10).

\begin{figure}[!htbp]
\centerline{\includegraphics[scale=0.75 ]{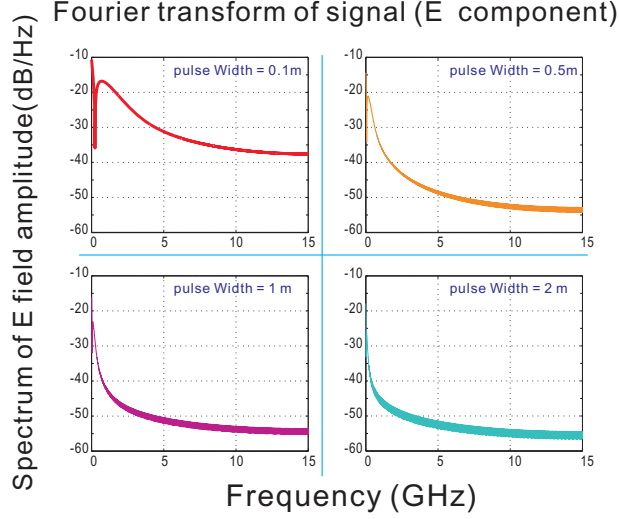}}
\begin{spacing}{1.2}
\caption{\footnotesize{\textbf{Amplitude spectrums of perturbed EM fields caused by the GW pulses with different widths.}
In the four subplots involving amplitude spectrums of perturbed EM fields, the corresponding  GW
pulses have different widths of $0.1~m$, $0.5~m$, $1~m$ and $2~m$ respectively(all amplitudes are here $\sim10^{-21}$).
It's manifestly obvious that the GW pulse with  smaller width (such as $0.1~m$), will definitely result in much more energy
distributed in the high frequency bands of the perturbed EM fields. Also, inversely, the GW pulse with larger width(such as $2m$), will lead to conditions for observing a very dramatic  attenuation of power of the perturbed EM fields in the high frequency bands.  The unit ``dB'' here means $10\times log_{10}( )$.}}
\end{spacing}
\end{figure}

\begin{figure}[!htbp]
\centerline{\includegraphics[scale=0.55]{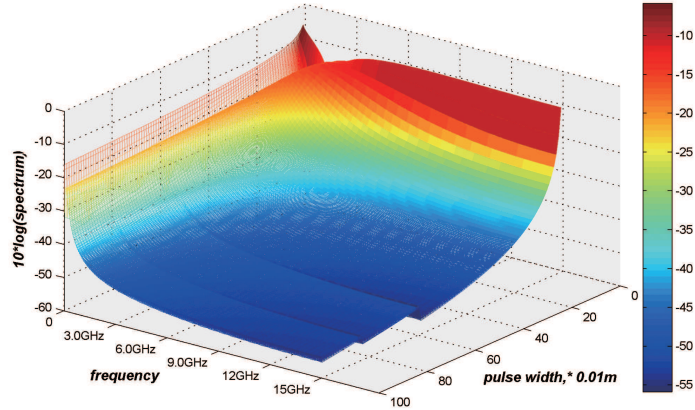}}
\begin{spacing}{1.2}
\caption{\footnotesize{\textbf{Continuous contours of amplitude spectrums of the perturbed EM fields, caused by GW with different frequency and pulse width.} Background magnetic field is set to $B=10^{11}$Tesla, and the given amplitude of the GW is set to $10^{-21}$. It's obvious that the GW pulse with smaller width will result in richer distribution of energy in high frequency bands(the area in red color.)}}
\end{spacing}
\end{figure}
~\\
\indent To summarize, based on the above interesting properties, we can argue that: the amplitude of
the GW pulse will proportionally influence the level and overall amplitude spectrum of the given
perturbed EM field, but nonlinearly, a smaller width of the GW pulse gives rise to a narrower
width of the EM pulse with higher strength of the peak, and smaller width brings greater proportion
of the energy distributed among the high-frequency band (e.g. GHz) in the spectrum.
In particular, we find that the perturbed EM pulses propagate at the speed of  light  synchronously with the GW pulse , leading to the accumulation  effect of their interaction, and it results in growing strength($\propto\sqrt{distance}$) of the perturbed EM fields within the region
of background magnetic field. These
important properties which connect perturbed impulsive EM fields and corresponding GW pulses
 supports the hypothesis that the obtained solutions are self-consistent and that they reasonably express the inherent physical nature of the EM response to cylindrical impulsive GW. Clearly, the
 magnetic component (Eq.(23)) of the perturbed EM field has similar properties.

\section{Electromagnetic response to the GW pulse by celestial or cosmological background magnetic fields and its potentially observable effects.}

The analytical solutions Eqs.(\ref{E}) and (\ref{H}) of  perturbed EM fields provide helpful information for studying the CSs,   impulsive cylindrical \mbox{GWs}, and relevant potentially \mbox{observable} effects. According to  classical electrodynamics, the power flux at a receiving surface $\Delta s$ of the perturbed EM fields may be expressed as:
\begin{equation}
\label{eq2}
 U =\frac{1}{2\mu_0}Re[{{E^{_{(1)}*}_y}(x, t)}\cdot{{B^{_{(1)}}_z}(x, t)}]
\Delta s
\end{equation}

The observability of the perturbed EM fields
will be determined concurrently by a lot of parameters of both GW pulse and other observation condition,
such as the amplitude  and  width
of the GW pulse, the strength of background magnetic field, the accumulation length (distance from GW source to receiving surface), the detecting technique for weak photons, the noise issue(and so on). Under current technology condition, the \mbox{detectable} minimal
EM power would be $\sim10^{-22}W$ in one Hz bandwidth\cite{Cruise_CQG29_2012},
so we could approximately assume the detectable minimal EM power for our case in this paper is the same order of magnitude. Also, the power of the perturbed EM fields is too week by only using current laboratory magnetic field (e.g., strength $\sim20~Tesla$,
accumulation distance $=3~m$ and area of receiving surface as $\Delta s=1.2 m^2$. Note that such signal will be much less than the minimal detectable EM power of $10^{-22}Watt$ ).  So, for obtaining the power of perturbed EM fields no less than this given minimal detectable level, we would need a very strong background magnetic field (e.g., say a celestial high magnetic field, which could reach up to $10^{11}$ Tesla\cite{Metzger.AstroPhysJourn.659.561.2007}) or some weak magnetic field but with extremely large scale (e.g. galactic-extragalactic background magnetic field\cite{galacticB.RevModPhys.74.775}, which leads to significant spatial accumulation effect) are required. These
together may permit detection via instrumentation.\\

\indent In Table.I,  it denotes the condition having background magnetic fields that are generated by some celestial bodies, such as neutron stars typically. These   astrophysical environments could act as natural laboratories. Contemporary researches believe that some young neutron stars can generate extremely high surface magnetic fields of $\sim 10^{10}$ to $ 10^{11} Tesla$\cite{Metzger.AstroPhysJourn.659.561.2007}.  Nevertheless, so far, our knowledge of exact parameters of the \mbox{GWs} from CSs,  is still relatively crude (including their amplitude, pulse width,  interval between adjacent pulses, and CSs' positions, distribution, spatial scale, etc.), but in keeping with previous estimation, the GWs from CSs could have amplitude $\sim10^{-31}$ or less (in high frequency range) in the Earth's region\cite{Abbott_Nature_2009}. If we study a specific CS, assuming the amplitude of GW emitted by it also has the same order of magnitude ($\sim10^{-31}$ or less) around the globe, then amplitude of the GW in the region near axis would be roughly $\sim10^{-21}$ (since the energy flux of cylindrical GW decays by $1/distance$ due to $\tau_0^1 \propto \frac{1}{x}$, see Eq.(\ref{tau01}), so the amplitude decays by $1/\sqrt{distance}$) provided that a possible source of CS would locate somewhere within the Galaxy (e.g., around center of the Galaxy, about 3000 light years or $\sim10^{19}m$ away from the globe); or, the amplitude of the GW in the region near axis would be roughly $\sim10^{-20}$ provided that the CS would locate around 1Mpc ($\sim10^{22}m$) away.  So under this circumstance, if there's very high magnetic fields (e.g., some from neutron stars or magnetars) also close to the CS(e.g., around the center of the Galaxy), then the EM response would lead to quite strong signal with power even up to $10^{2}~Watt$ (see Table.I case 1), that largely surpasses the minimal detectable EM power of $\sim10^{-22}Watt$. However, the method to measure these signals around the magnetars distant from the Earth is still immature, so it would only provide some considerable indirect effect.\\
\begin{table}[!htbp]
\begin{center} \centerline{
\begin{tabular}{|>{\small}c|>{\small}c|>{\small}c|>{\small}c|}
\hline
\multicolumn{4}{|c|}{~}\\
\multicolumn{4}{|c|}{Celestial condition: $ {B^{_{(0)}}_{SI}} =10^{11}~Tesla$, $\Delta s=100~m^2$,}\\
\multicolumn{4}{|c|}{accumulation distance$=1000~m$.}\\
\hline
~&~&~&~\\
case&signal power, &amplitude h~&width a\\
No.&&~of the GW pulse~~& of the GW pulse\\
~&~&~&~\\
\hline
~&~&~&~\\
1&~$1.15\times10^{~2}~~W~$	&~$10^{-21}$~&	0.1~m\\	
2&~$1.05\times10^{-1}~W~$	&~$10^{-21}$~&	~~1~m\\	
3&~$7.78\times10^{-5}~W~$	&~$10^{-21}$~&  10~m\\	
4&~$1.05\times10^{-3}~W~$	&~$10^{-22}$~&	~~1~m\\	
5&~$1.15\times10^{-2}~W~$	&~$10^{-23}$~&	0.1~m\\	
\hline
\end{tabular}}
\end{center}
\begin{spacing}{1.3}
\caption{ \textbf{Celestial  condition: typical potential ranges of power of the perturbed EM signals, with parameters of amplitude ``h'' and width ``a'' of the GW pulses.}
 Considering the perturbed EM signals interaction between impulsive cylindrical GWs with extremely high magnetic fields($\sim10^{11}Tesla$) produced by some celestial bodies such as neutron stars\cite{Metzger.AstroPhysJourn.659.561.2007} (here we denote $ {B^{_{(0)}}_{SI}} $ as background magnetic field in SI units), with the accumulation distance $\sim 1000~m$ and area of receiving surface $\Delta s\sim 100~m^2$ in the region near axis. }
\end{spacing}
\end{table}
\indent On the other hand,  as direct observable effect, the observation of expected perturbed EM fields even on the Earth is also possible.  The very widely existing  background galactic and extragalactic magnetic fields appearing in all galaxies and galaxy clusters\cite{galacticB.RevModPhys.74.775}, could give significant contribution to the spatial accumulation effect, during propagating of the GW pulse from its source to the Earth. So, taking this point into consideration, even if the CSs are very far from us, this galactic-extragalactic background magnetic fields (strength could reach $\sim10^{-9}$ Tesla within 1Mpc\cite{galacticB.RevModPhys.74.775}) will interact with the GW pulses in a huge accumulative distance. Then it would lead to observable signals in the Earth's region, because the accumulation effect of the perturbed EM fields is proportional to $\sqrt{distance}$ asymptotically($\sim\sqrt{x}$, see Eq.(\ref{Eaccumu}) and Fig.(6)). \\
\begin{figure}[!htbp]
\centerline{\includegraphics[scale=0.48]{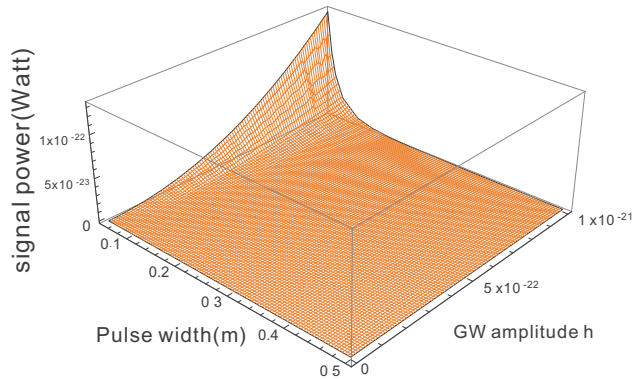}}
\begin{spacing}{1.2}
\caption{\footnotesize{\textbf{Continuous 3D plot of the power of perturbed EM fields (in far axis region, around the Earth), by different pulse width and amplitude of the GWs.} Here, the background galactic magnetic field is set to $10^{-9}$Tesla, accumulation distance is $2.8\times10^{19}m$, area of receiving surface is $5m^{2}$.}}
\end{spacing}
\end{figure}
\indent For example, as a rough estimation, if we set accumulation distance $\sim2.8\times10^{19}m$ (e.g. around center of the Galaxy), GW amplitude $h\sim 10^{-21}$ in the region near axis, receiving surface $\Delta s\sim 5 m^2$, GW pulse width $a\sim 5~cm$, galactic-extragalactic background magnetic field $B_{GMF}\sim10^{-9}~Tesla$, then the power of signal in the Earth's region would reach up to $1.3\times10^{-22} Watt$ (see Fig.11, it already surpasses the minimal detectable EM power of $10^{-22}Watt$, and it means that the photon flux is approximately 112 photons per second in GHz band). Besides, if this weak signal can be amplified by schemes of EM coupling\cite{FYLi_PRD67_2003,FYLi_EPJC_2008,34}, then higher signal power would be expected, and the parameters of $h$, $\Delta s$, $a$ and   $B_{GMF}$,  would also be enormously relaxed. Still, to extract the amplified signal from the EM coupling
system remains a very  challenging problem.
Moreover, if we consider a larger accumulation distance around 1Mpc $\sim 10^{22}m$, with other parameters of $h\sim 10^{-20}$, $\Delta s\sim 1 m^2$,  $a\sim 0.1~m$, $B_{GMF}\sim10^{-9}~Tesla$\cite{galacticB.RevModPhys.74.775}, then the power of signal on the Earth would reach up to a magnitude of $1.2\times10^{-19} Watt$ (which means that the photon flux is about $4.4\times10^4$ photons per second in the GHz band), which comes with greater direct observability.  This EM response caused by background galactic-extragalactic magnetic fields in all galaxies and galaxy clusters, would also be supplementary to the indirect observation of the \mbox{GWs} interacting with the cosmic microwave background (CMB) radiation\cite{Baskaran_PRD083008_2006,Polnarev_MNRAS1053_2008,Seljak_PRL2054_1997,
Pritchard_AnnPhysNY2_2005,Zhao_PRD083006_2006}  or some other effects.\\

\section{Asymptotic Behaviors}

Several representative asymptotic behaviors of the electric component ${E^{_{(1)}}_y}(x, t)$  of \mbox{perturbed} EM fields  (Eq.(\ref{E})) can be analytically deduced in diverse conditions, by which way the physical meaning of obtained solution would be expressed  more explicitly and simply. In this section we also demonstrate asymptotic behaviors of the energy density, energy flux density and Riemann curvature tensor of the impulsive cylindrical GW. The self-consistency, commonalities and differences among these asymptotic behaviors of both GW pulse and perturbed EM fields will be figured out. For convenience, we use natural units in the whole of this section.\\

\indent\textbf{(1)Asymptotic behavior of perturbed EM fields in space-like infinite region.}\\
 \indent When the EM pulses propagate in the area where $x\gg t$, and $~x\gg a$(width), from Eq.(\ref{E}) we have:
\begin{equation}
 ~~~~{E^{_{(1)}}_y}(x, t)\rightarrow A {B^{_{(0)}}}
 [\frac{4t}{x^2}-\frac{12a^2t}{x^4}] \propto ~~4A {B^{_{(0)}}} \frac{t}{x^2};
 \end{equation}
This asymptotic behavior shows that, at a specific time $t$,
${E^{_{(1)}}_y}(x, t)$ is weakening fast with respect to the term $1/distance^2$ along the $x$-axis.\\
\indent \textbf{(2)Asymptotic behavior of perturbed EM fields in time-like infinite region.}\\
 \indent When the EM pulses propagate in the area where $x\ll t$ , and $t\gg a$, also from Eq.(\ref{E}) we have:
 \begin{equation}
 ~~~~{E^{_{(1)}}_y}(x, t)\rightarrow A {B^{_{(0)}}}
 \frac{4x}{t^2}-4A  {B^{_{(0)}}} a^2\frac{1}{t^3} \propto ~~4A {B^{_{(0)}}} \frac{x}{t^2};
 \end{equation}
This asymptotic behavior of
${E^{(1)}_y}(x, t)$  indicates that, given a specific position $x$, EM pulses will fade away by the term $1/t^2$.\\
\indent\textbf{(3)Asymptotic behavior of perturbed EM fields in light-like infinite region} (on the light cone, i.e. $x=t\gg a$). \\
 \indent As shown in Fig.12, we define the region having background
  magnetic field $B^{(0)}$ is $x\geqslant x_0$, where $x_0$ is the position of the first
contact between wavefront of impulsive GW and $B^{(0)}$, then the interaction duration is from the start time $t_{min}=0$, to the end time $t_{max}=x-x_0=\Delta x$. In this case, Eq.(\ref{E}) approaches the form below (notice there's a factor c of the light speed in the electric component, but c=1 in the natural units here):
\begin{eqnarray}
  \label{Eaccumu}
~~~~~{E^{(1)}_y}(x, t)\rightarrow\frac{A {B^{(0)}} (a^4+4a^2\Delta x^2)^{\frac{1}{4}}}{a^2}  \nonumber\\
 \thickapprox \frac{A {B^{(0)}} [4a^2(x-x_0)^2]^{\frac{1}{4}}}{a^2}\propto \sqrt{x}
 \end{eqnarray}
\begin{figure}[H]
\centerline{\includegraphics[scale=0.9]{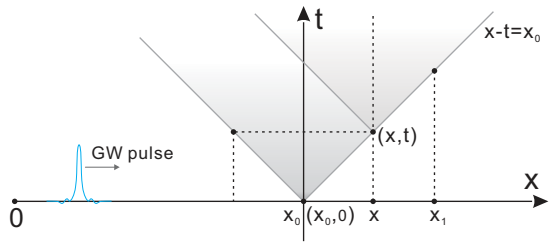}}
\begin{spacing}{1.2}
\caption{\footnotesize{\textbf{Asymptotic behavior of the perturbed EM fields in light-like region (on the light cone).}
}}
\end{spacing}
\end{figure}
Contrary to those E field in space-like or time-like infinite region, asymptotically  the ${E^{_{(1)}}_y}(x, t)$ on the light cone will not decay, but will increase, and this is particularly reflecting the known accumulation effect of the EM response (see Fig.6 and Sec.III). It indicates that the light cone is the most interesting area to observe the perturbed EM fields.  The magnetic component ${B^{_{(1)}}_z}(x, t)$ of the perturbed EM fields has similar asymptotic behaviors.\\

\indent\textbf{(4)Asymptotic behaviors of energy density and energy flux density of the impulsive cylindrical \mbox{GW}.}\\
 \indent The expression of energy density of impulsive cylindrical GW is\cite{FYLi.ActaPhysSin.321.1997}:
\begin{eqnarray}
\tau_0^0&=&\frac{1}{8\pi}e^{2\gamma}(\psi_x^2+\psi_t^2)\nonumber\\
&=& \frac{A^2}{2\pi}\{ \frac{x^2\cos^2\frac{3}{2}\theta+(t\cos\frac{3}{2}\theta-a\sin\frac{3}{2}\theta)^2}{[x^4+2x^2(a^2-t^2)+(a^2+t^2)^2]^{3/2})}\}\nonumber\\
&\cdot&\exp\{A^2[\frac{1}{a^2}-\frac{2x^2\cos2\theta}{ x^4+2x^2(a^2-t^2)+(a^2+t^2)^2  }\nonumber\\
&-&\frac{t^2+a^2-x^2}{a^2[t^4+2t^2(a^2-t^2)+(a^2+t^2)^2]^{1/2}}]\}
\end{eqnarray}
The expression of energy flux density of impulsive cylindrical GW is\cite{FYLi.ActaPhysSin.321.1997}:
\begin{eqnarray}
\tau_0^1 &=& -\frac{1}{4\pi}\psi_\rho\psi_t e^{2\gamma}\nonumber\\
&=& \frac{A^2}{2\pi}\{ \frac{x(2t\cos^2(\frac{3}{2}\theta)-a\sin(3\theta)}{[x^4+2x^2(a^2-t^2)+(a^2+t^2)^2]^{3/2})}\}\nonumber\\
&\cdot&\exp\{A^2[\frac{1}{a^2}-\frac{2x^2\cos2\theta}{ x^4+2x^2(a^2-t^2)+(a^2+t^2)^2  }\nonumber\\&-&\frac{t^2+a^2-x^2}{a^2[t^4+2t^2(a^2-t^2)+(a^2+t^2)^2]^{1/2}}]\}\nonumber\\
&where& , \theta=\theta_1= \arctan\frac{2at}{x^2+a^2-t^2}
\end{eqnarray}
\indent So we find that,\\
\indent (i) in space-like infinite region, where $x \gg t$, and  $x \gg a$,
we have the asymptotic behaviors of the energy density and energy flux density of the impulsive cylindrical GW as:
\begin{eqnarray}
&\tau&_0^0\rightarrow\frac{A^2}{2\pi  }(\frac{1}{x^4}+\frac{t^2}{x^6})\exp(\frac{2A^2}{a^2})\rightarrow O(x^{-4}),\nonumber\\
~\nonumber\\
&\tau&_0^1\rightarrow\frac{A^2t}{\pi x^5}\exp(\frac{2A^2}{a^2})\rightarrow O(x^{-5}),
\end{eqnarray}
where, energy density and energy flux density fall very quickly as the distance $x$ rises.\\
\indent(ii) in time-like infinite region, where $t \gg x$, and  $t \gg a$,
we have the asymptotic behavior of energy density and energy flux density as:
\begin{eqnarray}
&\tau&_0^0\rightarrow\frac{A^2}{2\pi t^4}\exp(-\frac{2A^2x^2}{t^4})\rightarrow O(t^{-4}),\nonumber\\
~\nonumber\\
&\tau&_0^1\rightarrow\frac{A^2x}{\pi t^5}\exp(-\frac{2A^2x^2}{t^4})\rightarrow O(t^{-5}),
\end{eqnarray}
where, energy density and energy flux density also drop rapidly as the time increases.\\
\indent (iii) in light-like infinite region (on the light cone, where $x=t\gg a$, as the most physically interesting region), their asymptotic behaviors are:
\begin{eqnarray}
&(&\tau_0^0)_{x=t\gg a}\rightarrow \frac{A^2}{16\pi a^3 x} \exp(\frac{3A^2}{2a^2})\propto \frac{1}{x},\nonumber\\
~\nonumber\\
\label{tau01}
&(&\tau_0^1)_{x=t\gg a}\rightarrow \frac{A^2}{16\pi a^3 x} \exp(\frac{3A^2}{2a^2})\propto \frac{1}{x}.
\end{eqnarray}
Therefore, the energy density and energy flux density here decay much slower on the light cone, comparing to those in space-like or time-like infinite regions. These asymptotic behaviors are consistent to  asymptotic behaviors of the perturbed EM fields shown above.\\

\indent\textbf{(5)Asymptotic behaviors of the Riemann curvature tensor of the impulsive cylindrical GW.}
We chose two typical non-vanishing components $R_{xzxz}$ and  $R_{y0y0}$  of covariant curvature tensor, and they have the forms\cite{FYLi.ActaPhysSin.321.1997}:
\begin{eqnarray}
&R&_{xzxz}\nonumber\\
&=&e^{2\psi}(-\psi_t\gamma_t+\psi_t^2-\psi_x\gamma_x+\psi_{xx})\nonumber\\
&=&e^{2\psi}\{\frac{-2A\cos\frac{3}{2}\theta}{r^{3/4}}+\frac{6Ax^2\cos\frac{5}{2}\theta}{r^{5/4}}\nonumber\\
&+&\frac{16A^3x^2\cos\frac{3}{2}\theta[x^2\cos^2\frac{3}{2}\theta+2(x\cos\frac{3}{2}\theta-a\sin\frac{3}{2}\theta)^2]}{r^{9/4}}\nonumber\\
&+&\frac{4A^2[2x^2\cos^2\frac{3}{2}\theta+(x\cos\frac{3}{2}\theta-a\sin\frac{3}{2}\theta)^2]}{r^{3/2}}\},
\end{eqnarray}
where $r=x^4+2x^2(a^2-t^2)+(a^2+t^2)^2$, the same hereinafter, and\\
\begin{eqnarray}
&R&_{y0y0}\nonumber\\
&=&-e^{-2\psi}(-\psi_{tt}-\psi_t\gamma_t+(\gamma_x-\psi_x)/x-\psi_x\gamma_x+\psi_x^2)\nonumber\\
&=&-e^{-2\psi}\{\frac{4A\cos\frac{3}{2}\theta}{r^{3/4}}+\frac{6A[(t^2-a^2)\cos\frac{5}{2}\theta-at\sin\frac{5}{2}\theta]}{r^{5/4}}\nonumber\\
&-&\frac{8A^3x^2\cos^2\frac{3}{2}\theta[x^2\cos^2\frac{3}{2}\theta-(t\cos\frac{3}{2}\theta-a\sin\frac{3}{2}\theta)^2]}{r^{9/4}}\nonumber\\
&+&\frac{4A^2[2x^2\cos^2\frac{3}{2}\theta+(t\cos\frac{3}{2}\theta-a\sin\frac{3}{2}\theta)^2]}{r^{3/2}}\},
\end{eqnarray}
\indent then we find that, \\
\indent (i) in space-like infinite region where $x\gg t$ and $x\gg a$, asymptotically it gives:
\begin{eqnarray}
R_{xzxz}&\rightarrow&(\frac{4A}{x^3}+\frac{12A^2}{x^4}+\frac{48A^3}{x^5})\exp(\frac{4A}{x})\nonumber\\
~\nonumber\\
&\rightarrow& \frac{4A}{x^3}\exp(\frac{4A}{x})\rightarrow O(\frac{1}{x^3}),
\end{eqnarray}
and
\begin{eqnarray}
R_{y0y0}&\rightarrow&-[\frac{4A}{x^3}+\frac{8A^2}{x^4}+\frac{6A(t^2-a^2)-8A^3}{x^5}]\exp(\frac{-4A}{x})\nonumber\\
~\nonumber\\
&\rightarrow& \frac{-4A}{x^3}\exp(\frac{-4A}{x})\rightarrow O(\frac{1}{x^3}),
\end{eqnarray}
\indent (ii) in time-like infinite region where $t\gg x$ and $t\gg a$, we have the asymptotic behaviors as:
\begin{eqnarray}
R_{xzxz}&\rightarrow&(\frac{-2A}{t^3}+\frac{6Ax^2}{t^5}+\frac{12A^2x^2}{t^6}+\frac{48A^3x^4}{t^9})\exp(\frac{4A}{t})\nonumber\\
~\nonumber\\
&\rightarrow& \frac{-2A}{t^3}\exp(\frac{4A}{t})\rightarrow O(\frac{1}{t^3}),
\end{eqnarray}
and
\begin{eqnarray}
R_{y0y0}&\rightarrow&(\frac{10A}{t^3}+\frac{4A^2}{t^4}+\frac{8A^3x^2}{t^7})\exp(\frac{-4A}{t})\nonumber\\
~\nonumber\\
&\rightarrow& \frac{-10A}{t^3}\exp(\frac{-4A}{t})\rightarrow O(\frac{1}{t^3}),
\end{eqnarray}
\indent (iii) in light-like infinite region (on the light cone), where \mbox{$x=t\gg a$}, their asymptotic behaviors are:
\begin{eqnarray}
R_{xzxz}&\rightarrow& -\frac{3}{4}(\frac{A}{a^{5/2}x^{1/2}}+\frac{A^3}{a^{9/2}x^{1/2}})\exp(\frac{-2A}{a^{1/2}x^{1/2}})\nonumber\\
~\nonumber\\
&\propto& \frac{1}{x^{1/2}}.
\end{eqnarray}
and
\begin{eqnarray}
R_{y0y0}&\rightarrow&(\frac{3A}{4a^{5/2}x^{1/2}}+\frac{3A^2}{4a^3x}+\frac{A}{a^{3/2}x^{3/2}})\exp(\frac{2A}{a^{1/2}x^{1/2}})\nonumber\\
~\nonumber\\
&\rightarrow&\frac{3A}{4a^{5/2}x^{1/2}}\exp(\frac{2A}{a^{1/2}x^{1/2}})\propto\frac{1}{x^{1/2}}\nonumber\\
\end{eqnarray}
Apparently, these components of Riemann curvature tensor \mbox{decline} much slowly on the light cone (the most interesting and concerned region with richest observable information), as compared to those in space-like or time-like infinite regions where they attenuate very rapidly. This characteristic agrees well with asymptotic behaviors of the energy density, energy flux density of the impulsive cylindrical GW, and asymptotic behaviors of the perturbed EM fields. Particularly, only on the light cone, perturbed EM fields will be growing instead of \mbox{declining}, that specially reflecting the spatial accumulation effect. All of these of above asymptotic
behaviors play supporting roles in further corroboration of the self-consistency and reasonability
of the obtained solutions.\\

\section{Concluding and discussing}
\indent First, in the frame of General Relativity, based upon  electrodynamical equations in curved spacetime, utilizing the d'Alembert Formula and relevant approaches, the analytical solutions $E^{(1)}_y(x, t)$ and $B^{(1)}_z(x, t)$ of the impulsive EM fields, perturbed by cylindrical GW pulses (could be emitted from cosmic strings) propagating through
background magnetic field, are obtained.
It's shown that the perturbed EM fields are also in the impulsive form, consistent with the impulsive cylindrical GWs, and the solutions can naturally give the accumulation effect  of  perturbed EM signal (due to the fact that perturbed EM pulses propagate at the speed of light synchronously with the propagating GW pulse), by the term of the square root of accumulated distance, i.e. $\propto\sqrt{distance}$. Based on this accumulation effect, we for the first time predict possible directly observable effect($\geq10^{-22}Watt$, stronger than the minimal detectable EM power under current experimental condition) on the Earth caused by the EM response of the GWs(from CSs) interacting with  background galactic-extragalactic magnetic fields.\\

\indent Second, asymptotic behaviors of the perturbed EM fields are accordant to asymptotic behaviors of the GW pulse and some of its relevant physical quantities such as energy density, energy flux density and Riemann curvature tensor, and it brings cogent affirmation supporting the self-consistency and reasonability of the obtained solutions.  Asymptotically, almost all of these  physical quantities will decline once the distance grows, and these physical quantities decline much more slowly on the light cone (in the light-like region) which is the most interesting area with the richest physical information, rather than the asymptotic behaviors
in the space-like or time-like regions where they attenuate rapidly. Whereas, only perturbed
EM fields in the light cone will not decline, but instead, increase. Also, we find that the asymptotic behaviors of perturbed EM fields
particularly reflect the profile and dynamical behavior of the spatial accumulation effect.\\

\indent   Third, perturbed EM fields caused by the cylindrical impulsive \mbox{GWs} from CSs are often very weak, and then direct detection or indirect observation would be very difficult on the Earth.
However, many contemporary research results convince us that there are extremely high magnetic fields in some celestial bodies' regions (such as neutron stars\cite{Metzger.AstroPhysJourn.659.561.2007}, which could cause indirect observable effect), and very widely distributed  galactic-extragalactic background magnetic fields in all galaxies and galaxy clusters\cite{galacticB.RevModPhys.74.775}; and especially the latter might provide a huge spatial accumulation effect for the perturbed EM fields, and  would lead to very interesting and potentially observable effect in the Earth's region (as the effect for direct observation, see section V), even if such CSs are distant from the Earth (e.g., locate around  center of the Galaxy, i.e., about 3000 light years away, or even further, like $\sim$1Mpc).\\
\indent In addition we find analysis of representative physical properties of the perturbed EM fields also
reveals that: (1)amplitude of GW pulse proportionally influences the level and overall spectrum of
the perturbed EM field. (2)Smaller width of the GW pulse nonlinearly gives rise to  narrower
widths, higher peaks of the perturbed EM pulses, and greater proportion of energy distributed in
the  high-frequency band (e.g. GHz) in the amplitude spectrums of perturbed EM fields.\\
\indent According to previous studies, the cylindrical \mbox{GWs} from CSs include both impulsive and usual continuous forms.
Specially, in this paper we only focus on the impulsive case due to its concentrated energy, the pre-existing rigorous metric (\mbox{Einstein}-\mbox{Rosen} metric), and its impulsive property to give rich GW \mbox{components} covering wide frequency band, etc. Nevertheless, EM response to the usual continuous \mbox{GWs}, also would bring meaningful information and value for in-depth studying in future. Moreover, in order to enhance the real detectability of the perturbed EM fields, various considerable improvements could be introduced, such as EM coupling\cite{FYLi_PRD67_2003,FYLi_EPJC_2008,9,8} and also use of superconductivity cavity technology\cite{34}. We intend to have through investigations of these additional research topics in the future.\\

\begin{acknowledgments}
This work is supported by the National Nature \mbox{Science} Foundation of China No.11375279,
the Foundation of China Academy of Engineering Physics No.2008 T0401 and T0402.
\end{acknowledgments}

\bibliographystyle{apsrev4-1}
\bibliography{WenReferenceData}

\providecommand{\noopsort}[1]{}\providecommand{\singleletter}[1]{#1}%
\begin{thebibliography}{61}%
\makeatletter
\providecommand \@ifxundefined [1]{%
 \@ifx{#1\undefined}
}%
\providecommand \@ifnum [1]{%
 \ifnum #1\expandafter \@firstoftwo
 \else \expandafter \@secondoftwo
 \fi
}%
\providecommand \@ifx [1]{%
 \ifx #1\expandafter \@firstoftwo
 \else \expandafter \@secondoftwo
 \fi
}%
\providecommand \natexlab [1]{#1}%
\providecommand \enquote  [1]{``#1''}%
\providecommand \bibnamefont  [1]{#1}%
\providecommand \bibfnamefont [1]{#1}%
\providecommand \citenamefont [1]{#1}%
\providecommand \href@noop [0]{\@secondoftwo}%
\providecommand \href [0]{\begingroup \@sanitize@url \@href}%
\providecommand \@href[1]{\@@startlink{#1}\@@href}%
\providecommand \@@href[1]{\endgroup#1\@@endlink}%
\providecommand \@sanitize@url [0]{\catcode `\\12\catcode `\$12\catcode
  `\&12\catcode `\#12\catcode `\^12\catcode `\_12\catcode `\%12\relax}%
\providecommand \@@startlink[1]{}%
\providecommand \@@endlink[0]{}%
\providecommand \url  [0]{\begingroup\@sanitize@url \@url }%
\providecommand \@url [1]{\endgroup\@href {#1}{\urlprefix }}%
\providecommand \urlprefix  [0]{URL }%
\providecommand \Eprint [0]{\href }%
\providecommand \doibase [0]{http://dx.doi.org/}%
\providecommand \selectlanguage [0]{\@gobble}%
\providecommand \bibinfo  [0]{\@secondoftwo}%
\providecommand \bibfield  [0]{\@secondoftwo}%
\providecommand \translation [1]{[#1]}%
\providecommand \BibitemOpen [0]{}%
\providecommand \bibitemStop [0]{}%
\providecommand \bibitemNoStop [0]{.\EOS\space}%
\providecommand \EOS [0]{\spacefactor3000\relax}%
\providecommand \BibitemShut  [1]{\csname bibitem#1\endcsname}%
\let\auto@bib@innerbib\@empty
\bibitem [{\citenamefont {BICEP2.Collaboration}()}]{B.mode.CMB}%
  \BibitemOpen
  \bibfield  {author} {\bibinfo {author} {\bibnamefont
  {BICEP2.Collaboration}},\ }\href@noop {} {\bibinfo  {journal}
  {arXiv:1403.3985v2}\ }\BibitemShut {NoStop}%
\bibitem [{\citenamefont {Allen}()}]{Allen_arXiv9604033}%
  \BibitemOpen
\bibfield  {journal} {  }\bibfield  {author} {\bibinfo {author} {\bibfnamefont
  {B.}~\bibnamefont {Allen}},\ }\href@noop {} {\bibinfo  {journal}
  {arXiv:gr-qc/9604033}\ }\BibitemShut {NoStop}%
\bibitem [{\citenamefont {Vilenkin}(1981)}]{Vilenkin_PLB47_1981}%
  \BibitemOpen
\bibfield  {journal} {  }\bibfield  {author} {\bibinfo {author} {\bibfnamefont
  {A.}~\bibnamefont {Vilenkin}},\ }\href@noop {} {\bibfield  {journal}
  {\bibinfo  {journal} {Phys. Lett. B}\ }\textbf {\bibinfo {volume} {107}},\
  \bibinfo {pages} {47} (\bibinfo {year} {1981})}\BibitemShut {NoStop}%
\bibitem [{\citenamefont {Caldwell}\ and\ \citenamefont
  {Allen}(1992)}]{Caldwell_PRD3447_1992}%
  \BibitemOpen
  \bibfield  {author} {\bibinfo {author} {\bibfnamefont {R.~R.}\ \bibnamefont
  {Caldwell}}\ and\ \bibinfo {author} {\bibfnamefont {B.}~\bibnamefont
  {Allen}},\ }\href@noop {} {\bibfield  {journal} {\bibinfo  {journal} {Phys.
  Rev. D}\ }\textbf {\bibinfo {volume} {45}},\ \bibinfo {pages} {3447}
  (\bibinfo {year} {1992})}\BibitemShut {NoStop}%
\bibitem [{\citenamefont {Vachaspati}\ and\ \citenamefont
  {Vilenkin}(1985)}]{Vachaspati_PhysRevD.31.3052}%
  \BibitemOpen
  \bibfield  {author} {\bibinfo {author} {\bibfnamefont {T.}~\bibnamefont
  {Vachaspati}}\ and\ \bibinfo {author} {\bibfnamefont {A.}~\bibnamefont
  {Vilenkin}},\ }\href {\doibase 10.1103/PhysRevD.31.3052} {\bibfield
  {journal} {\bibinfo  {journal} {Phys. Rev. D}\ }\textbf {\bibinfo {volume}
  {31}},\ \bibinfo {pages} {3052} (\bibinfo {year} {1985})}\BibitemShut
  {NoStop}%
\bibitem [{\citenamefont {Hogan}\ and\ \citenamefont
  {Rees}(1984)}]{HOGAN_Nature_1984}%
  \BibitemOpen
  \bibfield  {author} {\bibinfo {author} {\bibfnamefont {C.~J.}\ \bibnamefont
  {Hogan}}\ and\ \bibinfo {author} {\bibfnamefont {M.~J.}\ \bibnamefont
  {Rees}},\ }\href@noop {} {\bibfield  {journal} {\bibinfo  {journal} {Nature}\
  }\textbf {\bibinfo {volume} {311}},\ \bibinfo {pages} {109} (\bibinfo {year}
  {1984})}\BibitemShut {NoStop}%
\bibitem [{\citenamefont {Damour}\ and\ \citenamefont
  {Vilenkin}(2000)}]{Damour_PRL3761_2000}%
  \BibitemOpen
  \bibfield  {author} {\bibinfo {author} {\bibfnamefont {T.}~\bibnamefont
  {Damour}}\ and\ \bibinfo {author} {\bibfnamefont {A.}~\bibnamefont
  {Vilenkin}},\ }\href@noop {} {\bibfield  {journal} {\bibinfo  {journal}
  {Phys. Rev. Lett.}\ }\textbf {\bibinfo {volume} {85}},\ \bibinfo {pages}
  {3761} (\bibinfo {year} {2000})}\BibitemShut {NoStop}%
\bibitem [{\citenamefont {Damour}\ and\ \citenamefont
  {Vilenkin}(2005)}]{Damour_PRD063510_2005}%
  \BibitemOpen
  \bibfield  {author} {\bibinfo {author} {\bibfnamefont {T.}~\bibnamefont
  {Damour}}\ and\ \bibinfo {author} {\bibfnamefont {A.}~\bibnamefont
  {Vilenkin}},\ }\href@noop {} {\bibfield  {journal} {\bibinfo  {journal}
  {Phys. Rev. D}\ }\textbf {\bibinfo {volume} {71}},\ \bibinfo {pages} {063510}
  (\bibinfo {year} {2005})}\BibitemShut {NoStop}%
\bibitem [{\citenamefont {Leblond}\ \emph {et~al.}(2009)\citenamefont
  {Leblond}, \citenamefont {Shlaer},\ and\ \citenamefont
  {Siemens}}]{Leblond_PRD123519_2009}%
  \BibitemOpen
  \bibfield  {author} {\bibinfo {author} {\bibfnamefont {L.}~\bibnamefont
  {Leblond}}, \bibinfo {author} {\bibfnamefont {B.}~\bibnamefont {Shlaer}}, \
  and\ \bibinfo {author} {\bibfnamefont {X.}~\bibnamefont {Siemens}},\
  }\href@noop {} {\bibfield  {journal} {\bibinfo  {journal} {Phys. Rev. D}\
  }\textbf {\bibinfo {volume} {79}},\ \bibinfo {pages} {123519} (\bibinfo
  {year} {2009})}\BibitemShut {NoStop}%
\bibitem [{\citenamefont {Dufaux}\ \emph {et~al.}(2010)\citenamefont {Dufaux},
  \citenamefont {Figueroa},\ and\ \citenamefont
  {Garc{\'i}a-Bellido}}]{Dufaux_PRD083518_2010}%
  \BibitemOpen
  \bibfield  {author} {\bibinfo {author} {\bibfnamefont {J.~F.}\ \bibnamefont
  {Dufaux}}, \bibinfo {author} {\bibfnamefont {D.~G.}\ \bibnamefont
  {Figueroa}}, \ and\ \bibinfo {author} {\bibfnamefont {J.}~\bibnamefont
  {Garc{\'i}a-Bellido}},\ }\href {\doibase 10.1103/PhysRevD.82.083518}
  {\bibfield  {journal} {\bibinfo  {journal} {Phys. Rev. D}\ }\textbf {\bibinfo
  {volume} {82}},\ \bibinfo {pages} {083518} (\bibinfo {year}
  {2010})}\BibitemShut {NoStop}%
\bibitem [{\citenamefont {Berezinsky}\ \emph {et~al.}(2001)\citenamefont
  {Berezinsky}, \citenamefont {Hnatyk},\ and\ \citenamefont
  {Vilenkin}}]{Berezinsky_PRD043004_2001}%
  \BibitemOpen
  \bibfield  {author} {\bibinfo {author} {\bibfnamefont {V.}~\bibnamefont
  {Berezinsky}}, \bibinfo {author} {\bibfnamefont {B.}~\bibnamefont {Hnatyk}},
  \ and\ \bibinfo {author} {\bibfnamefont {A.}~\bibnamefont {Vilenkin}},\
  }\href@noop {} {\bibfield  {journal} {\bibinfo  {journal} {Phys. Rev. D}\
  }\textbf {\bibinfo {volume} {64}},\ \bibinfo {pages} {043004} (\bibinfo
  {year} {2001})}\BibitemShut {NoStop}%
\bibitem [{\citenamefont {Copeland}\ \emph {et~al.}(2004)\citenamefont
  {Copeland}, \citenamefont {Myers},\ and\ \citenamefont
  {Polchinski}}]{Copeland_JHEP013_2004}%
  \BibitemOpen
  \bibfield  {author} {\bibinfo {author} {\bibfnamefont {E.~J.}\ \bibnamefont
  {Copeland}}, \bibinfo {author} {\bibfnamefont {R.~C.}\ \bibnamefont {Myers}},
  \ and\ \bibinfo {author} {\bibfnamefont {J.}~\bibnamefont {Polchinski}},\
  }\href@noop {} {\bibfield  {journal} {\bibinfo  {journal} {J. High Energy
  Phys.}\ }\textbf {\bibinfo {volume} {06}},\ \bibinfo {pages} {013} (\bibinfo
  {year} {2004})}\BibitemShut {NoStop}%
\bibitem [{\citenamefont {Siemens}\ \emph {et~al.}(2006)\citenamefont
  {Siemens}, \citenamefont {Creighton}, \citenamefont {Maor}, \citenamefont
  {Majumder}, \citenamefont {Cannon},\ and\ \citenamefont
  {Read}}]{Siemens.PRD.105001}%
  \BibitemOpen
  \bibfield  {author} {\bibinfo {author} {\bibfnamefont {X.}~\bibnamefont
  {Siemens}}, \bibinfo {author} {\bibfnamefont {J.}~\bibnamefont {Creighton}},
  \bibinfo {author} {\bibfnamefont {I.}~\bibnamefont {Maor}}, \bibinfo {author}
  {\bibfnamefont {S.~R.}\ \bibnamefont {Majumder}}, \bibinfo {author}
  {\bibfnamefont {K.}~\bibnamefont {Cannon}}, \ and\ \bibinfo {author}
  {\bibfnamefont {J.}~\bibnamefont {Read}},\ }\href {\doibase
  10.1103/PhysRevD.73.105001} {\bibfield  {journal} {\bibinfo  {journal} {Phys.
  Rev. D}\ }\textbf {\bibinfo {volume} {73}},\ \bibinfo {pages} {105001}
  (\bibinfo {year} {2006})}\BibitemShut {NoStop}%
\bibitem [{\citenamefont {Podolsk{\'y}}\ and\ \citenamefont
  {Griffiths}(2000)}]{Podolsky.CQG.1401.2000}%
  \BibitemOpen
  \bibfield  {author} {\bibinfo {author} {\bibfnamefont {J.}~\bibnamefont
  {Podolsk{\'y}}}\ and\ \bibinfo {author} {\bibfnamefont {J.~B.}\ \bibnamefont
  {Griffiths}},\ }\href@noop {} {\bibfield  {journal} {\bibinfo  {journal}
  {Class. Quantum Grav.}\ }\textbf {\bibinfo {volume} {17}},\ \bibinfo {pages}
  {1401} (\bibinfo {year} {2000})}\BibitemShut {NoStop}%
\bibitem [{\citenamefont {Podolsk\'y}\ and\ \citenamefont
  {\ifmmode~\check{S}\else \v{S}\fi{}varc}(2010)}]{Podolsk.PhysRevD.81.124035}%
  \BibitemOpen
  \bibfield  {author} {\bibinfo {author} {\bibfnamefont {J.}~\bibnamefont
  {Podolsk\'y}}\ and\ \bibinfo {author} {\bibfnamefont {R.}~\bibnamefont
  {\ifmmode~\check{S}\else \v{S}\fi{}varc}},\ }\href {\doibase
  10.1103/PhysRevD.81.124035} {\bibfield  {journal} {\bibinfo  {journal} {Phys.
  Rev. D}\ }\textbf {\bibinfo {volume} {81}},\ \bibinfo {pages} {124035}
  (\bibinfo {year} {2010})}\BibitemShut {NoStop}%
\bibitem [{\citenamefont {Gleiser}\ and\ \citenamefont
  {Pullin}(1989)}]{Gleiser.CQG.L141.1989}%
  \BibitemOpen
  \bibfield  {author} {\bibinfo {author} {\bibfnamefont {R.}~\bibnamefont
  {Gleiser}}\ and\ \bibinfo {author} {\bibfnamefont {J.}~\bibnamefont
  {Pullin}},\ }\href@noop {} {\bibfield  {journal} {\bibinfo  {journal} {Class.
  Quantum Grav.}\ }\textbf {\bibinfo {volume} {6}},\ \bibinfo {pages} {L141}
  (\bibinfo {year} {1989})}\BibitemShut {NoStop}%
\bibitem [{\citenamefont {Slagter}(2001)}]{Slagter.CQG.463.2001}%
  \BibitemOpen
  \bibfield  {author} {\bibinfo {author} {\bibfnamefont {R.~J.}\ \bibnamefont
  {Slagter}},\ }\href@noop {} {\bibfield  {journal} {\bibinfo  {journal}
  {Class. Quantum Grav.}\ }\textbf {\bibinfo {volume} {18}},\ \bibinfo {pages}
  {463} (\bibinfo {year} {2001})}\BibitemShut {NoStop}%
\bibitem [{\citenamefont {Horta\ifmmode~\mbox{\c{c}}\else
  \c{c}\fi{}su}(1996)}]{Hortacsu.CQG.2683}%
  \BibitemOpen
  \bibfield  {author} {\bibinfo {author} {\bibfnamefont {M.}~\bibnamefont
  {Horta\ifmmode~\mbox{\c{c}}\else \c{c}\fi{}su}},\ }\href
  {http://stacks.iop.org/0264-9381/13/i=10/a=008} {\bibfield  {journal}
  {\bibinfo  {journal} {Classical and Quantum Gravity}\ }\textbf {\bibinfo
  {volume} {13}},\ \bibinfo {pages} {2683} (\bibinfo {year}
  {1996})}\BibitemShut {NoStop}%
\bibitem [{\citenamefont {Steinbauer}\ and\ \citenamefont
  {Vickers}(2006)}]{Steinbauer.CQG.2006}%
  \BibitemOpen
  \bibfield  {author} {\bibinfo {author} {\bibfnamefont {R.}~\bibnamefont
  {Steinbauer}}\ and\ \bibinfo {author} {\bibfnamefont {J.~A.}\ \bibnamefont
  {Vickers}},\ }\href {http://stacks.iop.org/0264-9381/23/i=10/a=R01}
  {\bibfield  {journal} {\bibinfo  {journal} {Classical and Quantum Gravity}\
  }\textbf {\bibinfo {volume} {23}},\ \bibinfo {pages} {R91} (\bibinfo {year}
  {2006})}\BibitemShut {NoStop}%
\bibitem [{\citenamefont {Dubath}\ and\ \citenamefont
  {Rocha}(2007)}]{Dubath.PRD.024001}%
  \BibitemOpen
  \bibfield  {author} {\bibinfo {author} {\bibfnamefont {F.}~\bibnamefont
  {Dubath}}\ and\ \bibinfo {author} {\bibfnamefont {J.~V.}\ \bibnamefont
  {Rocha}},\ }\href {\doibase 10.1103/PhysRevD.76.024001} {\bibfield  {journal}
  {\bibinfo  {journal} {Phys. Rev. D}\ }\textbf {\bibinfo {volume} {76}},\
  \bibinfo {pages} {024001} (\bibinfo {year} {2007})}\BibitemShut {NoStop}%
\bibitem [{\citenamefont {{\"O}lmez}\ \emph {et~al.}(2010)\citenamefont
  {{\"O}lmez}, \citenamefont {Mandic},\ and\ \citenamefont
  {Siemens}}]{Olmez_PhysRevD.81.104028}%
  \BibitemOpen
  \bibfield  {author} {\bibinfo {author} {\bibfnamefont {S.}~\bibnamefont
  {{\"O}lmez}}, \bibinfo {author} {\bibfnamefont {V.}~\bibnamefont {Mandic}}, \
  and\ \bibinfo {author} {\bibfnamefont {X.}~\bibnamefont {Siemens}},\ }\href
  {\doibase 10.1103/PhysRevD.81.104028} {\bibfield  {journal} {\bibinfo
  {journal} {Phys. Rev. D}\ }\textbf {\bibinfo {volume} {81}},\ \bibinfo
  {pages} {104028} (\bibinfo {year} {2010})}\BibitemShut {NoStop}%
\bibitem [{\citenamefont {Patel}\ \emph {et~al.}(2010)\citenamefont {Patel},
  \citenamefont {Siemens}, \citenamefont {Dupuis},\ and\ \citenamefont
  {Betzwieser}}]{Patel.PRD.084032}%
  \BibitemOpen
  \bibfield  {author} {\bibinfo {author} {\bibfnamefont {P.}~\bibnamefont
  {Patel}}, \bibinfo {author} {\bibfnamefont {X.}~\bibnamefont {Siemens}},
  \bibinfo {author} {\bibfnamefont {R.}~\bibnamefont {Dupuis}}, \ and\ \bibinfo
  {author} {\bibfnamefont {J.}~\bibnamefont {Betzwieser}},\ }\href {\doibase
  10.1103/PhysRevD.81.084032} {\bibfield  {journal} {\bibinfo  {journal} {Phys.
  Rev. D}\ }\textbf {\bibinfo {volume} {81}},\ \bibinfo {pages} {084032}
  (\bibinfo {year} {2010})}\BibitemShut {NoStop}%
\bibitem [{\citenamefont {Kleidis}\ \emph {et~al.}(2010)\citenamefont
  {Kleidis}, \citenamefont {Kuiroukidis}, \citenamefont {Nerantzi},\ and\
  \citenamefont {Papadopoulos}}]{Kleidis.2010.31}%
  \BibitemOpen
  \bibfield  {author} {\bibinfo {author} {\bibfnamefont {K.}~\bibnamefont
  {Kleidis}}, \bibinfo {author} {\bibfnamefont {A.}~\bibnamefont
  {Kuiroukidis}}, \bibinfo {author} {\bibfnamefont {P.}~\bibnamefont
  {Nerantzi}}, \ and\ \bibinfo {author} {\bibfnamefont {D.}~\bibnamefont
  {Papadopoulos}},\ }\href {\doibase 10.1007/s10714-009-0812-z} {\bibfield
  {journal} {\bibinfo  {journal} {General Relativity and Gravitation}\ }\textbf
  {\bibinfo {volume} {42}},\ \bibinfo {pages} {31} (\bibinfo {year}
  {2010})}\BibitemShut {NoStop}%
\bibitem [{\citenamefont {Hindmarsh}\ and\ \citenamefont {{T. B. W.
  Kibble}}(1995)}]{Hindmarsh_RepProgrPhys447_1995}%
  \BibitemOpen
  \bibfield  {author} {\bibinfo {author} {\bibfnamefont {M.~B.}\ \bibnamefont
  {Hindmarsh}}\ and\ \bibinfo {author} {\bibnamefont {{T. B. W. Kibble}}},\
  }\href@noop {} {\bibfield  {journal} {\bibinfo  {journal} {Rep. Progr.
  Phys.}\ }\textbf {\bibinfo {volume} {58}},\ \bibinfo {pages} {477} (\bibinfo
  {year} {1995})}\BibitemShut {NoStop}%
\bibitem [{\citenamefont {Vilenkin}\ and\ \citenamefont {{E. P. S.
  Shellard}}(2000)}]{Vilenkin_Cambridge_2000}%
  \BibitemOpen
  \bibfield  {author} {\bibinfo {author} {\bibfnamefont {A.}~\bibnamefont
  {Vilenkin}}\ and\ \bibinfo {author} {\bibnamefont {{E. P. S. Shellard}}},\
  }\href@noop {} {\emph {\bibinfo {title} {Cosmic Strings and Other Topological
  Defects}}}\ (\bibinfo  {publisher} {Cambridge University Press},\ \bibinfo
  {address} {Cambridge},\ \bibinfo {year} {2000})\BibitemShut {NoStop}%
\bibitem [{\citenamefont {Wang}\ and\ \citenamefont
  {Santos}(1996)}]{AnzhongWang.CQG.715}%
  \BibitemOpen
  \bibfield  {author} {\bibinfo {author} {\bibfnamefont {A.}~\bibnamefont
  {Wang}}\ and\ \bibinfo {author} {\bibfnamefont {N.~O.}\ \bibnamefont
  {Santos}},\ }\href {http://stacks.iop.org/0264-9381/13/i=4/a=011} {\bibfield
  {journal} {\bibinfo  {journal} {Classical and Quantum Gravity}\ }\textbf
  {\bibinfo {volume} {13}},\ \bibinfo {pages} {715} (\bibinfo {year}
  {1996})}\BibitemShut {NoStop}%
\bibitem [{\citenamefont {Gregory}(1989)}]{Gregory.PhysRevD.39.2108}%
  \BibitemOpen
  \bibfield  {author} {\bibinfo {author} {\bibfnamefont {R.}~\bibnamefont
  {Gregory}},\ }\href {\doibase 10.1103/PhysRevD.39.2108} {\bibfield  {journal}
  {\bibinfo  {journal} {Phys. Rev. D}\ }\textbf {\bibinfo {volume} {39}},\
  \bibinfo {pages} {2108} (\bibinfo {year} {1989})}\BibitemShut {NoStop}%
\bibitem [{\citenamefont {Abbott}\ \emph
  {et~al.}(2009{\natexlab{a}})\citenamefont {Abbott} \emph
  {et~al.}}]{Abbott_PRD062002_2009}%
  \BibitemOpen
  \bibfield  {author} {\bibinfo {author} {\bibfnamefont {B.~P.}\ \bibnamefont
  {Abbott}} \emph {et~al.},\ }\href@noop {} {\bibfield  {journal} {\bibinfo
  {journal} {Phys. Rev. D}\ }\textbf {\bibinfo {volume} {80}},\ \bibinfo
  {pages} {062002} (\bibinfo {year} {2009}{\natexlab{a}})}\BibitemShut
  {NoStop}%
\bibitem [{\citenamefont {Siemens}\ \emph {et~al.}(2007)\citenamefont
  {Siemens}, \citenamefont {Mandic},\ and\ \citenamefont
  {Creighton}}]{Siemens_PhysRevLett.98.111101}%
  \BibitemOpen
  \bibfield  {author} {\bibinfo {author} {\bibfnamefont {X.}~\bibnamefont
  {Siemens}}, \bibinfo {author} {\bibfnamefont {V.}~\bibnamefont {Mandic}}, \
  and\ \bibinfo {author} {\bibfnamefont {J.}~\bibnamefont {Creighton}},\ }\href
  {\doibase 10.1103/PhysRevLett.98.111101} {\bibfield  {journal} {\bibinfo
  {journal} {Phys. Rev. Lett.}\ }\textbf {\bibinfo {volume} {98}},\ \bibinfo
  {pages} {111101} (\bibinfo {year} {2007})}\BibitemShut {NoStop}%
\bibitem [{\citenamefont {Bin{\'e}truy}\ \emph {et~al.}(2010)\citenamefont
  {Bin{\'e}truy}, \citenamefont {Boh{\'e}}, \citenamefont {Hertog},\ and\
  \citenamefont {Steer}}]{Binetruy_PhysRevD.82.126007}%
  \BibitemOpen
  \bibfield  {author} {\bibinfo {author} {\bibfnamefont {P.~P.}\ \bibnamefont
  {Bin{\'e}truy}}, \bibinfo {author} {\bibfnamefont {A.}~\bibnamefont
  {Boh{\'e}}}, \bibinfo {author} {\bibfnamefont {T.}~\bibnamefont {Hertog}}, \
  and\ \bibinfo {author} {\bibfnamefont {D.~A.}\ \bibnamefont {Steer}},\ }\href
  {\doibase 10.1103/PhysRevD.82.126007} {\bibfield  {journal} {\bibinfo
  {journal} {Phys. Rev. D}\ }\textbf {\bibinfo {volume} {82}},\ \bibinfo
  {pages} {126007} (\bibinfo {year} {2010})}\BibitemShut {NoStop}%
\bibitem [{\citenamefont {Cohen}\ \emph {et~al.}(2010)\citenamefont {Cohen},
  \citenamefont {Cutler},\ and\ \citenamefont
  {Vallisneri}}]{Cohen_CQG185012_2010}%
  \BibitemOpen
  \bibfield  {author} {\bibinfo {author} {\bibfnamefont {M.~I.}\ \bibnamefont
  {Cohen}}, \bibinfo {author} {\bibfnamefont {C.}~\bibnamefont {Cutler}}, \
  and\ \bibinfo {author} {\bibfnamefont {M.}~\bibnamefont {Vallisneri}},\
  }\href@noop {} {\bibfield  {journal} {\bibinfo  {journal} {Class. Quantum
  Grav.}\ }\textbf {\bibinfo {volume} {27}},\ \bibinfo {pages} {185012}
  (\bibinfo {year} {2010})}\BibitemShut {NoStop}%
\bibitem [{\citenamefont {{E. O'Callaghan}}\ \emph {et~al.}(2010)\citenamefont
  {{E. O'Callaghan}}, \citenamefont {Chadburn}, \citenamefont {Geshnizjani},
  \citenamefont {Gregory},\ and\ \citenamefont
  {Zavala}}]{Callaghan_PhysRevLett.105.081602}%
  \BibitemOpen
  \bibfield  {author} {\bibinfo {author} {\bibnamefont {{E. O'Callaghan}}},
  \bibinfo {author} {\bibfnamefont {S.}~\bibnamefont {Chadburn}}, \bibinfo
  {author} {\bibfnamefont {G.}~\bibnamefont {Geshnizjani}}, \bibinfo {author}
  {\bibfnamefont {R.}~\bibnamefont {Gregory}}, \ and\ \bibinfo {author}
  {\bibfnamefont {I.}~\bibnamefont {Zavala}},\ }\href {\doibase
  10.1103/PhysRevLett.105.081602} {\bibfield  {journal} {\bibinfo  {journal}
  {Phys. Rev. Lett.}\ }\textbf {\bibinfo {volume} {105}},\ \bibinfo {pages}
  {081602} (\bibinfo {year} {2010})}\BibitemShut {NoStop}%
\bibitem [{\citenamefont {Bennett}\ and\ \citenamefont
  {Bouchet}(1988)}]{Bennett_PhysRevLett.60.257}%
  \BibitemOpen
  \bibfield  {author} {\bibinfo {author} {\bibfnamefont {D.~P.}\ \bibnamefont
  {Bennett}}\ and\ \bibinfo {author} {\bibfnamefont {F.~R.}\ \bibnamefont
  {Bouchet}},\ }\href {\doibase 10.1103/PhysRevLett.60.257} {\bibfield
  {journal} {\bibinfo  {journal} {Phys. Rev. Lett.}\ }\textbf {\bibinfo
  {volume} {60}},\ \bibinfo {pages} {257} (\bibinfo {year} {1988})}\BibitemShut
  {NoStop}%
\bibitem [{\citenamefont {Caldwell}\ \emph {et~al.}(1996)\citenamefont
  {Caldwell}, \citenamefont {Battye},\ and\ \citenamefont
  {Shellard}}]{Caldwell_PhysRevD.54.7146}%
  \BibitemOpen
  \bibfield  {author} {\bibinfo {author} {\bibfnamefont {R.~R.}\ \bibnamefont
  {Caldwell}}, \bibinfo {author} {\bibfnamefont {R.~A.}\ \bibnamefont
  {Battye}}, \ and\ \bibinfo {author} {\bibfnamefont {E.~P.~S.}\ \bibnamefont
  {Shellard}},\ }\href {\doibase 10.1103/PhysRevD.54.7146} {\bibfield
  {journal} {\bibinfo  {journal} {Phys. Rev. D}\ }\textbf {\bibinfo {volume}
  {54}},\ \bibinfo {pages} {7146} (\bibinfo {year} {1996})}\BibitemShut
  {NoStop}%
\bibitem [{\citenamefont {Sarangi}\ and\ \citenamefont
  {Tye}(2002)}]{Sarangi.PLB185.2002}%
  \BibitemOpen
  \bibfield  {author} {\bibinfo {author} {\bibfnamefont {S.}~\bibnamefont
  {Sarangi}}\ and\ \bibinfo {author} {\bibfnamefont {S.}~\bibnamefont {Tye}},\
  }\href@noop {} {\bibfield  {journal} {\bibinfo  {journal} {Phys. Lett. B}\
  }\textbf {\bibinfo {volume} {536}},\ \bibinfo {pages} {185} (\bibinfo {year}
  {2002})}\BibitemShut {NoStop}%
\bibitem [{\citenamefont {Li}\ \emph {et~al.}(2003)\citenamefont {Li},
  \citenamefont {Tang},\ and\ \citenamefont {Shi}}]{FYLi_PRD67_2003}%
  \BibitemOpen
  \bibfield  {author} {\bibinfo {author} {\bibfnamefont {F.~Y.}\ \bibnamefont
  {Li}}, \bibinfo {author} {\bibfnamefont {M.~X.}\ \bibnamefont {Tang}}, \ and\
  \bibinfo {author} {\bibfnamefont {D.~P.}\ \bibnamefont {Shi}},\ }\href
  {\doibase 10.1103/PhysRevD.67.104008} {\bibfield  {journal} {\bibinfo
  {journal} {Phys. Rev. D}\ }\textbf {\bibinfo {volume} {67}},\ \bibinfo
  {pages} {104008} (\bibinfo {year} {2003})}\BibitemShut {NoStop}%
\bibitem [{\citenamefont {Li}\ \emph {et~al.}(2008)\citenamefont {Li},
  \citenamefont {{R. M. L. Baker, Jr.}}, \citenamefont {Fang}, \citenamefont
  {Stepheson},\ and\ \citenamefont {Chen}}]{FYLi_EPJC_2008}%
  \BibitemOpen
  \bibfield  {author} {\bibinfo {author} {\bibfnamefont {F.~Y.}\ \bibnamefont
  {Li}}, \bibinfo {author} {\bibnamefont {{R. M. L. Baker, Jr.}}}, \bibinfo
  {author} {\bibfnamefont {Z.~Y.}\ \bibnamefont {Fang}}, \bibinfo {author}
  {\bibfnamefont {G.~V.}\ \bibnamefont {Stepheson}}, \ and\ \bibinfo {author}
  {\bibfnamefont {Z.~Y.}\ \bibnamefont {Chen}},\ }\href@noop {} {\bibfield
  {journal} {\bibinfo  {journal} {Eur. Phys. J. C}\ }\textbf {\bibinfo {volume}
  {56}},\ \bibinfo {pages} {407} (\bibinfo {year} {2008})}\BibitemShut
  {NoStop}%
\bibitem [{\citenamefont {Li}\ \emph {et~al.}(2009)\citenamefont {Li},
  \citenamefont {Yang}, \citenamefont {Fang}, \citenamefont {Baker},
  \citenamefont {Stephenson},\ and\ \citenamefont {Wen}}]{FYLi_PRD80_2009}%
  \BibitemOpen
  \bibfield  {author} {\bibinfo {author} {\bibfnamefont {F.}~\bibnamefont
  {Li}}, \bibinfo {author} {\bibfnamefont {N.}~\bibnamefont {Yang}}, \bibinfo
  {author} {\bibfnamefont {Z.}~\bibnamefont {Fang}}, \bibinfo {author}
  {\bibfnamefont {R.~M.~L.}\ \bibnamefont {Baker}}, \bibinfo {author}
  {\bibfnamefont {G.~V.}\ \bibnamefont {Stephenson}}, \ and\ \bibinfo {author}
  {\bibfnamefont {H.}~\bibnamefont {Wen}},\ }\href@noop {} {\bibfield
  {journal} {\bibinfo  {journal} {Phys. Rev. D}\ }\textbf {\bibinfo {volume}
  {80}},\ \bibinfo {pages} {064013} (\bibinfo {year} {2009})}\BibitemShut
  {NoStop}%
\bibitem [{\citenamefont {{W. K. De Logi}}\ and\ \citenamefont
  {Mickelson}(1977)}]{DeLogi_PRD16_1977}%
  \BibitemOpen
  \bibfield  {author} {\bibinfo {author} {\bibnamefont {{W. K. De Logi}}}\ and\
  \bibinfo {author} {\bibfnamefont {A.~R.}\ \bibnamefont {Mickelson}},\ }\href
  {\doibase 10.1103/PhysRevD.16.2915} {\bibfield  {journal} {\bibinfo
  {journal} {Phys. Rev. D}\ }\textbf {\bibinfo {volume} {16}},\ \bibinfo
  {pages} {2915} (\bibinfo {year} {1977})}\BibitemShut {NoStop}%
\bibitem [{\citenamefont {Boccaletti}\ \emph {et~al.}(1970)\citenamefont
  {Boccaletti}, \citenamefont {{V. De Sabbata}}, \citenamefont {Fortint},\ and\
  \citenamefont {Gualdi}}]{Boccaletti_NuovoCim70_1970}%
  \BibitemOpen
  \bibfield  {author} {\bibinfo {author} {\bibfnamefont {D.}~\bibnamefont
  {Boccaletti}}, \bibinfo {author} {\bibnamefont {{V. De Sabbata}}}, \bibinfo
  {author} {\bibfnamefont {P.}~\bibnamefont {Fortint}}, \ and\ \bibinfo
  {author} {\bibfnamefont {C.}~\bibnamefont {Gualdi}},\ }\href {\doibase
  10.1007/BF02710177} {\bibfield  {journal} {\bibinfo  {journal} {Nuovo Cim.
  B}\ }\textbf {\bibinfo {volume} {70}},\ \bibinfo {pages} {129} (\bibinfo
  {year} {1970})}\BibitemShut {NoStop}%
\bibitem [{\citenamefont {Einstein}\ and\ \citenamefont
  {Rosen}(1937)}]{Einstein.JFI.43.1937}%
  \BibitemOpen
  \bibfield  {author} {\bibinfo {author} {\bibfnamefont {A.}~\bibnamefont
  {Einstein}}\ and\ \bibinfo {author} {\bibfnamefont {N.}~\bibnamefont
  {Rosen}},\ }\href@noop {} {\bibfield  {journal} {\bibinfo  {journal} {J.
  Franklin Inst.}\ }\textbf {\bibinfo {volume} {223}},\ \bibinfo {pages} {43}
  (\bibinfo {year} {1937})}\BibitemShut {NoStop}%
\bibitem [{\citenamefont {Rosen}(1937)}]{Rosen.PhyZSowjet.366.1937}%
  \BibitemOpen
  \bibfield  {author} {\bibinfo {author} {\bibfnamefont {N.}~\bibnamefont
  {Rosen}},\ }\href@noop {} {\bibfield  {journal} {\bibinfo  {journal} {Physik
  Z. Sowjetunion}\ }\textbf {\bibinfo {volume} {12}},\ \bibinfo {pages} {366}
  (\bibinfo {year} {1937})}\BibitemShut {NoStop}%
\bibitem [{\citenamefont {Widrow}(2002)}]{galacticB.RevModPhys.74.775}%
  \BibitemOpen
  \bibfield  {author} {\bibinfo {author} {\bibfnamefont {L.~M.}\ \bibnamefont
  {Widrow}},\ }\href {\doibase 10.1103/RevModPhys.74.775} {\bibfield  {journal}
  {\bibinfo  {journal} {Rev. Mod. Phys.}\ }\textbf {\bibinfo {volume} {74}},\
  \bibinfo {pages} {775} (\bibinfo {year} {2002})}\BibitemShut {NoStop}%
\bibitem [{\citenamefont {Metzger}\ \emph {et~al.}(2007)\citenamefont
  {Metzger}, \citenamefont {Thompson},\ and\ \citenamefont
  {Quataert}}]{Metzger.AstroPhysJourn.659.561.2007}%
  \BibitemOpen
  \bibfield  {author} {\bibinfo {author} {\bibfnamefont {B.~D.}\ \bibnamefont
  {Metzger}}, \bibinfo {author} {\bibfnamefont {T.~A.}\ \bibnamefont
  {Thompson}}, \ and\ \bibinfo {author} {\bibfnamefont {E.}~\bibnamefont
  {Quataert}},\ }\href@noop {} {\bibfield  {journal} {\bibinfo  {journal}
  {Astrophys. J.}\ }\textbf {\bibinfo {volume} {659}},\ \bibinfo {pages} {561}
  (\bibinfo {year} {2007})}\BibitemShut {NoStop}%
\bibitem [{\citenamefont {Weber}(1961)}]{weber.1961.GRGW}%
  \BibitemOpen
  \bibfield  {author} {\bibinfo {author} {\bibfnamefont {J.}~\bibnamefont
  {Weber}},\ }\href {http://books.google.com.hk/books?id=t6\_AXrEnVZYC} {\emph
  {\bibinfo {title} {General Relativity And Gravitational Waves}}},\ Dover
  Books on Physics Series\ (\bibinfo  {publisher} {Dover Publications},\
  \bibinfo {year} {1961})\BibitemShut {NoStop}%
\bibitem [{\citenamefont {Weber}\ and\ \citenamefont
  {Wheeler}(1957)}]{Weber.RevModPhys.509.1957}%
  \BibitemOpen
  \bibfield  {author} {\bibinfo {author} {\bibfnamefont {J.}~\bibnamefont
  {Weber}}\ and\ \bibinfo {author} {\bibfnamefont {J.~A.}\ \bibnamefont
  {Wheeler}},\ }\href@noop {} {\bibfield  {journal} {\bibinfo  {journal} {Rev.
  Mod. Phys.}\ }\textbf {\bibinfo {volume} {29}},\ \bibinfo {pages} {509}
  (\bibinfo {year} {1957})}\BibitemShut {NoStop}%
\bibitem [{\citenamefont {Rosen}\ and\ \citenamefont
  {Virbhadra}(1993)}]{Rosen.GRG.429.1993}%
  \BibitemOpen
  \bibfield  {author} {\bibinfo {author} {\bibfnamefont {N.}~\bibnamefont
  {Rosen}}\ and\ \bibinfo {author} {\bibfnamefont {K.~S.}\ \bibnamefont
  {Virbhadra}},\ }\href@noop {} {\bibfield  {journal} {\bibinfo  {journal}
  {Gen. Rel. Grav.}\ }\textbf {\bibinfo {volume} {25}},\ \bibinfo {pages} {429}
  (\bibinfo {year} {1993})}\BibitemShut {NoStop}%
\bibitem [{\citenamefont {Rosen}(1956)}]{Rosen.HeivPhysActa.171.1956}%
  \BibitemOpen
  \bibfield  {author} {\bibinfo {author} {\bibfnamefont {N.}~\bibnamefont
  {Rosen}},\ }\href@noop {} {\bibfield  {journal} {\bibinfo  {journal} {Heiv.
  Phys. Acta}\ }\textbf {\bibinfo {volume} {Suppl IV}},\ \bibinfo {pages} {171}
  (\bibinfo {year} {1956})}\BibitemShut {NoStop}%
\bibitem [{\citenamefont {Rosen}(1958)}]{Rosen.PhysRev.291.1958}%
  \BibitemOpen
  \bibfield  {author} {\bibinfo {author} {\bibfnamefont {N.}~\bibnamefont
  {Rosen}},\ }\href@noop {} {\bibfield  {journal} {\bibinfo  {journal} {Phys.
  Rev.}\ }\textbf {\bibinfo {volume} {110}},\ \bibinfo {pages} {291} (\bibinfo
  {year} {1958})}\BibitemShut {NoStop}%
\bibitem [{\citenamefont {Li}\ and\ \citenamefont
  {Tang}(1997)}]{FYLi.ActaPhysSin.321.1997}%
  \BibitemOpen
  \bibfield  {author} {\bibinfo {author} {\bibfnamefont {F.~Y.}\ \bibnamefont
  {Li}}\ and\ \bibinfo {author} {\bibfnamefont {M.~X.}\ \bibnamefont {Tang}},\
  }\href@noop {} {\bibfield  {journal} {\bibinfo  {journal} {Acta Phys. Sin.}\
  }\textbf {\bibinfo {volume} {6}},\ \bibinfo {pages} {321} (\bibinfo {year}
  {1997})}\BibitemShut {NoStop}%
\bibitem [{\citenamefont {Tikhonov}\ and\ \citenamefont
  {Samarskii}(2011)}]{Tikhonov}%
  \BibitemOpen
  \bibfield  {author} {\bibinfo {author} {\bibfnamefont {A.~N.}\ \bibnamefont
  {Tikhonov}}\ and\ \bibinfo {author} {\bibfnamefont {A.~A.}\ \bibnamefont
  {Samarskii}},\ }in\ \href@noop {} {\emph {\bibinfo {booktitle} {Equations of
  Mathematical Physics}}},\ \bibinfo {series and number} {Dover Books on
  Physics},\ \bibinfo {editor} {edited by\ \bibinfo {editor} {\bibfnamefont
  {D.~M.}\ \bibnamefont {Brink}}}\ (\bibinfo  {publisher} {Dover publications,
  Inc.},\ \bibinfo {address} {New York},\ \bibinfo {year} {2011})\ Chap.\
  \bibinfo {chapter} {II-2-7}, p.~\bibinfo {pages} {73},\ \bibinfo {edition}
  {dover}\ ed.\BibitemShut {Stop}%
\bibitem [{\citenamefont {Cruise}(2012)}]{Cruise_CQG29_2012}%
  \BibitemOpen
  \bibfield  {author} {\bibinfo {author} {\bibfnamefont {A.~M.}\ \bibnamefont
  {Cruise}},\ }\href@noop {} {\bibfield  {journal} {\bibinfo  {journal} {Class.
  Quantum Grav.}\ }\textbf {\bibinfo {volume} {29}},\ \bibinfo {pages} {095003}
  (\bibinfo {year} {2012})}\BibitemShut {NoStop}%
\bibitem [{\citenamefont {Abbott}\ \emph
  {et~al.}(2009{\natexlab{b}})\citenamefont {Abbott} \emph
  {et~al.}}]{Abbott_Nature_2009}%
  \BibitemOpen
  \bibfield  {author} {\bibinfo {author} {\bibfnamefont {B.~P.}\ \bibnamefont
  {Abbott}} \emph {et~al.},\ }\href {\doibase 10.1038/nature08278} {\bibfield
  {journal} {\bibinfo  {journal} {Nature (London)}\ }\textbf {\bibinfo {volume}
  {460}},\ \bibinfo {pages} {990} (\bibinfo {year}
  {2009}{\natexlab{b}})}\BibitemShut {NoStop}%
\bibitem [{\citenamefont {Li}\ \emph {et~al.}(2007)\citenamefont {Li},
  \citenamefont {Chen},\ and\ \citenamefont {Wang}}]{34}%
  \BibitemOpen
  \bibfield  {author} {\bibinfo {author} {\bibfnamefont {F.~Y.}\ \bibnamefont
  {Li}}, \bibinfo {author} {\bibfnamefont {Y.}~\bibnamefont {Chen}}, \ and\
  \bibinfo {author} {\bibfnamefont {P.}~\bibnamefont {Wang}},\ }\href@noop {}
  {\bibfield  {journal} {\bibinfo  {journal} {Chin. Phys. Lett.}\ }\textbf
  {\bibinfo {volume} {24}},\ \bibinfo {pages} {3328} (\bibinfo {year}
  {2007})}\BibitemShut {NoStop}%
\bibitem [{\citenamefont {Baskaran}\ \emph {et~al.}(2006)\citenamefont
  {Baskaran}, \citenamefont {Grishchuk},\ and\ \citenamefont
  {Polnarev}}]{Baskaran_PRD083008_2006}%
  \BibitemOpen
  \bibfield  {author} {\bibinfo {author} {\bibfnamefont {D.}~\bibnamefont
  {Baskaran}}, \bibinfo {author} {\bibfnamefont {L.~P.}\ \bibnamefont
  {Grishchuk}}, \ and\ \bibinfo {author} {\bibfnamefont {A.~G.}\ \bibnamefont
  {Polnarev}},\ }\href@noop {} {\bibfield  {journal} {\bibinfo  {journal}
  {Phys. Rev. D}\ }\textbf {\bibinfo {volume} {74}},\ \bibinfo {pages} {083008}
  (\bibinfo {year} {2006})}\BibitemShut {NoStop}%
\bibitem [{\citenamefont {A.~G.~Polnarev}(2008)}]{Polnarev_MNRAS1053_2008}%
  \BibitemOpen
  \bibfield  {author} {\bibinfo {author} {\bibfnamefont {B.~G.~K.}\
  \bibnamefont {A.~G.~Polnarev}, \bibfnamefont {N.~J.~Miller}},\ }\href@noop {}
  {\bibfield  {journal} {\bibinfo  {journal} {Monthly Notices of the Royal
  Astronomical Society}\ }\textbf {\bibinfo {volume} {386}},\ \bibinfo {pages}
  {1053} (\bibinfo {year} {2008})}\BibitemShut {NoStop}%
\bibitem [{\citenamefont {Seljak}\ and\ \citenamefont
  {Zaldarriaga}(1997)}]{Seljak_PRL2054_1997}%
  \BibitemOpen
  \bibfield  {author} {\bibinfo {author} {\bibfnamefont {U.}~\bibnamefont
  {Seljak}}\ and\ \bibinfo {author} {\bibfnamefont {M.}~\bibnamefont
  {Zaldarriaga}},\ }\href@noop {} {\bibfield  {journal} {\bibinfo  {journal}
  {Phys. Rev. Lett.}\ }\textbf {\bibinfo {volume} {78}},\ \bibinfo {pages}
  {2054} (\bibinfo {year} {1997})}\BibitemShut {NoStop}%
\bibitem [{\citenamefont {Pritchard}\ and\ \citenamefont
  {Kamionkowski}(2005)}]{Pritchard_AnnPhysNY2_2005}%
  \BibitemOpen
  \bibfield  {author} {\bibinfo {author} {\bibfnamefont {J.~R.}\ \bibnamefont
  {Pritchard}}\ and\ \bibinfo {author} {\bibfnamefont {M.}~\bibnamefont
  {Kamionkowski}},\ }\href@noop {} {\bibfield  {journal} {\bibinfo  {journal}
  {Ann. Phys. (N.Y.)}\ }\textbf {\bibinfo {volume} {318}},\ \bibinfo {pages}
  {2} (\bibinfo {year} {2005})}\BibitemShut {NoStop}%
\bibitem [{\citenamefont {Zhao}\ and\ \citenamefont
  {Zhang}(2006)}]{Zhao_PRD083006_2006}%
  \BibitemOpen
  \bibfield  {author} {\bibinfo {author} {\bibfnamefont {W.}~\bibnamefont
  {Zhao}}\ and\ \bibinfo {author} {\bibfnamefont {Y.}~\bibnamefont {Zhang}},\
  }\href@noop {} {\bibfield  {journal} {\bibinfo  {journal} {Phys. Rev. D}\
  }\textbf {\bibinfo {volume} {74}},\ \bibinfo {pages} {083006} (\bibinfo
  {year} {2006})}\BibitemShut {NoStop}%
\bibitem [{\citenamefont {Li}\ \emph {et~al.}(2011)\citenamefont {Li},
  \citenamefont {Lin}, \citenamefont {Li},\ and\ \citenamefont {Zhong}}]{9}%
  \BibitemOpen
  \bibfield  {author} {\bibinfo {author} {\bibfnamefont {J.}~\bibnamefont
  {Li}}, \bibinfo {author} {\bibfnamefont {K.}~\bibnamefont {Lin}}, \bibinfo
  {author} {\bibfnamefont {F.~Y.}\ \bibnamefont {Li}}, \ and\ \bibinfo {author}
  {\bibfnamefont {Y.~H.}\ \bibnamefont {Zhong}},\ }\href@noop {} {\bibfield
  {journal} {\bibinfo  {journal} {Gen. Relativ. Gravit.}\ }\textbf {\bibinfo
  {volume} {43}},\ \bibinfo {pages} {2209} (\bibinfo {year}
  {2011})}\BibitemShut {NoStop}%
\bibitem [{\citenamefont {Woods}\ \emph {et~al.}(2011)\citenamefont {Woods}
  \emph {et~al.}}]{8}%
  \BibitemOpen
  \bibfield  {author} {\bibinfo {author} {\bibfnamefont {R.}~\bibnamefont
  {Woods}} \emph {et~al.},\ }\href@noop {} {\bibfield  {journal} {\bibinfo
  {journal} {J. Mod. Phys.}\ }\textbf {\bibinfo {volume} {2}},\ \bibinfo
  {pages} {498} (\bibinfo {year} {2011})}\BibitemShut {NoStop}%
\end{thebibliography}%
\end{document}